\documentclass{aa}
\usepackage{times}
\usepackage{amsmath}
\usepackage{graphicx}
\begin{document}

\newcommand{\Msun}{\ensuremath{\mathrm{M}_\odot}}
\newcommand{\Lsun}{\ensuremath{\mathrm{L}_\odot}}
\newcommand{\Rsun}{\ensuremath{\mathrm{R}_\odot}}
\newcommand{\Mdot}{\ensuremath{\dot{M}}}
\newcommand{\nabad}{\ensuremath{\nabla_{\rm ad}}}
\newcommand{\nab}{\ensuremath{\nabla}}
\newcommand{\nabmu}{\ensuremath{\nabla_{\mu}}}
\newcommand{\nvt}{\ensuremath{N^2 _{\rm T}}}
\newcommand{\nvmu}{\ensuremath{N^2 _{\mu}}}
\newcommand{\fnabmu}{\ensuremath{(\varphi_{\rm e}\nabla_{\mu_{\rm e}}+\varphi_{\rm I}\nabla_{\mu_{\rm I}})/\delta }}
\newcommand{\density}{\ensuremath{\mathrm{g/cm^3}}}
\newcommand{\arrow}{$\longrightarrow$}
\newcommand{\lsim}{\mathrel{\hbox{\rlap{\lower.55ex \hbox {$\sim$}}
 \kern-.3em \raise.4ex \hbox{$<$}}}}
\newcommand{\gsim}{\mathrel{\hbox{\rlap{\lower.55ex \hbox {$\sim$}}
 \kern-.3em \raise.4ex \hbox{$>$}}}}
\newcommand{\half}{\ensuremath{{\textstyle\frac{1}{2}}}}
\newcommand{\e}{\ensuremath{\mathrm{e}}}
\newcommand{\n}{\ensuremath{\mathrm{n}}}
\newcommand{\p}{\ensuremath{\mathrm{p}}}
              \renewcommand{\H}[1]{\ensuremath{{^{#1}\mathrm{H}}}}
\newcommand{\D}{\ensuremath{\mathrm{D}}}
\newcommand{\He}[1]{\ensuremath{{^{#1}\mathrm{He}}}}
\newcommand{\Li}[1]{\ensuremath{{^{#1}\mathrm{Li}}}}
\newcommand{\Be}[1]{\ensuremath{{^{#1}\mathrm{Be}}}}
\newcommand{\B}[1]{\ensuremath{{^{#1}\mathrm{B}}}}
\newcommand{\C}[1]{\ensuremath{{^{#1}\mathrm{C}}}}
\newcommand{\N}[1]{\ensuremath{{^{#1}\mathrm{N}}}}
\renewcommand{\O}[1]{\ensuremath{{^{#1}\mathrm{O}}}}
\newcommand{\F}[1]{\ensuremath{{^{#1}\mathrm{F}}}}
\newcommand{\Na}[1]{\ensuremath{{^{#1}\mathrm{Na}}}}
\newcommand{\Ne}[1]{\ensuremath{{^{#1}\mathrm{Ne}}}}
\newcommand{\Mg}[1]{\ensuremath{{^{#1}\mathrm{Mg}}}}
\newcommand{\Fe}[1]{\ensuremath{{^{#1}\mathrm{Fe}}}}
\newcommand{\Ni}[1]{\ensuremath{{^{#1}\mathrm{Ni}}}}
\newcommand{\Co}[1]{\ensuremath{{^{#1}\mathrm{Co}}}}
\newcommand{\Al}[1]{\ensuremath{{^{#1}\mathrm{Al}}}}
\newcommand{\avg}[1]{\ensuremath{\langle#1\rangle}}
\newcommand{\rate}[1]{\ensuremath{r_\mathrm{#1}}}
\newcommand{\sv}[1]{\ensuremath{\langle\sigma v\rangle_\mathrm{#1}}}
\newcommand{\life}[2]{\ensuremath{\tau_\mathrm{#1}(\mathrm{#2})}}
\newcommand{\fpp}[1]{\ensuremath{f_\mathrm{pp#1}}}
                                                                                
\newcommand{\Mwd}{$M_{\rm{WD}}$}
\newcommand{\Mch}{$M_{\rm{Ch}}$}
\newcommand{\Msn}{$M_{\rm{SN}}$}
\newcommand{\Minit}{$M_{\rm{init}}$}
\newcommand{\Mmax}{$M_{\rm{max}}$}
\newcommand{\Mr}{$M_{\rm{r}}$}
\newcommand{\Mrdsi}{$M_{\rm{r,DSI}}$}
\newcommand{\sigdsi}{$\sigma_{\rm{DSI, crit}}$}
\newcommand{\msyr}{$\rm{M}_{\odot}/\rm{yr}$}
\newcommand{\kmpersec}{$\rm{km}\rm{s}^{-1}$}
\newcommand{\ctw}{$^{12}\rm{C}$}
\newcommand{\ost}{$^{16}\rm{O}$}
\newcommand{\rtrip}{$3\alpha$}
\newcommand{\rche}{$^{12}\rm{C}(\alpha,\gamma)^{16}\rm{O}$}
\newcommand{\mdota}{\ensuremath{3\cdot10^{-7}}}
\newcommand{\mdotb}{\ensuremath{5\cdot10^{-7}}}
\newcommand{\mdotc}{\ensuremath{7\cdot10^{-7}}}
\newcommand{\mdotd}{\ensuremath{10^{-6}}}
\newcommand{\taur}{\ensuremath{\tau_{\rm r}}}
\newcommand{\taub}{\ensuremath{\tau_{\rm bar}}}
\newcommand{\taun}{\ensuremath{\tau_{\nu}}}
\newcommand{\taua}{\ensuremath{\tau_{\rm acc}}}
\newcommand{\Mtwc}{\ensuremath{M_{{\rm crit,}~E_{\rm rot}/|W|}}}
\newcommand{\Mcrit}[1]{\ensuremath{M_{{\rm crit,}~{#1}}}}
\newcommand{\TW}{\ensuremath{E_{\rm rot}/|W|}}
\newcommand{\TWc}{\ensuremath{(E_{\rm rot}/|W|)_{\rm bar,~crit}}}
\newcommand{\Erot}{\ensuremath{\dot{E}_{\rm rot}}}
\newcommand{\Vaisala}{\ensuremath{{\rm V\ddot{a}is\ddot{a}l\ddot{a}} }}

% \thesaurus{ 06 (08.02.7; 08.03.4; 08.05.3; 08.18.1; 08.19.5)}
\newcommand{\ms}{$\rm{M}_{\odot}$}
\authorrunning{S.-C. Yoon \& N.  Langer}
\titlerunning{Pre-supernova evolution of accreting WDs with rotation}

\title{
Presupernova evolution of accreting white dwarfs with rotation 
}

\author{S.-C. Yoon \and N. Langer}

\institute{Astronomical Institute, Utrecht University, Princetonplein 5,
NL-3584 CC, Utrecht, The Netherlands}

\offprints {S.-C. Yoon (email: {\tt S.C.Yoon@astro.uu.nl})}
\date{Received  ; accepted }

%%%%%%%%%%%%%%%%%%%%%%%%%%%%%%%%%%%%%%%%%%%%%%%%%%%%%%%%%%%%%%%%%%%%%%%%%%%%%%%%%%%%%%%%%%%%

\abstract{ 
We discuss the effects of rotation
on the evolution of accreting carbon-oxygen white dwarfs, with the emphasis on 
possible consequences in Type Ia supernova (SN Ia) progenitors.
Starting with a slowly rotating white dwarf, we consider the accretion of
matter and angular momentum from a quasi-Keplerian accretion disk.
Numerical simulations with initial white dwarf masses of 0.8, 0.9 and  1.0 \Msun{}
and accretion of carbon-oxygen rich matter at
rates of $3\dots10\times10^{-7}$ \msyr{} are performed.
The models are evolved either up to a ratio of rotational to potential energy of
$T/W=0.18$ --- as angular momentum loss
through gravitational wave radiation will become important for $T/W < 0.18$ --- 
or to central carbon ignition.
The role of the various rotationally induced hydrodynamic instabilities for
the transport of angular momentum inside the white dwarf is investigated. We find that
the dynamical shear instability is the most important one in
the highly degenerate core, while Eddington Sweet circulation, 
Goldreich-Schubert-Fricke instability and secular shear instability
are most relevant in the non-degenerate envelope. 
Our results imply that accreting white dwarfs rotate differentially throughout,
with a shear rate close to 
the threshold value for the onset of the dynamical shear instability. 
As the latter depends on the temperature of the white dwarf, the thermal evolution
of the white dwarf core is found to be relevant for the angular momentum redistribution.
As found previously, significant rotation is shown to lead to
carbon ignition masses well above 1.4~\Msun.
Our models suggest a wide range of white dwarf explosion masses,
which could be responsible for some aspects of
the diversity  observed in SNe~Ia.
We analyze the potential role of the bar-mode and the $r$-mode instability 
in rapidly rotating white dwarfs,  
which may impose angular momentum loss by gravitational wave radiation.
We discuss the consequences of the resulting spin-down for the fate of the
white dwarf, and the possibility to detect the emitted gravitational waves
at frequencies of $0.1 \dots 1.0$ Hz in nearby galaxies with LISA.
Possible implications of fast and differentially rotating white dwarf cores for 
the flame propagation in exploding white dwarfs are also briefly discussed.
  
}
\maketitle

%%%%%%%%%%%%%%%%%%%%%%%%%%%%%%%%%%%%%%%%%%%%%%%%%%%%%%%%%%%%%%%%%%%%%%%%%%%%%%%%%%%%%%%%%%%%

\section{Introduction}

Type Ia Supernovae (SNe Ia) have a particular importance in astrophysics. 
Observations of SNe Ia at low redshift showed
a clear correlation between the peak brightness
and the width of the light curve (Phillips~\cite{Phillips93}; Phillips' relation), which allowed
to use SNe Ia as distance indicators for galaxies even beyond $z=1$. 
This made SNe Ia an indispensable tool for cosmology, in particular
to determine the cosmological parameters  
(e.g. Hamuy et al.~\cite{Hamuy96}; Branch~\cite{Branch98}; Leibundgut~\cite{Leibundgut00},~\cite{Leibundgut01}).
The new cosmology with a non-zero cosmological
constant has been initiated by 
the observational evidence deduced from observations of SNe Ia at high 
redshift (Perlmutter et al.~\cite{Perlmutter99}; Riess et al.~\cite{Riess00}).

Recent analyses of SNe Ia have revealed, however, that
SNe Ia are not perfectly homogeneous but
show some diversity in their light curves and spectra 
(e.g. Branch~\cite{Branch01}; Nomoto et al.~\cite{Nomoto03}; Li et al.~\cite{Li03}). 
This leaves concerns about applying the Phillips' relation to very distant SNe Ia.
An understanding of the origin of the diversity observed in SNe Ia
is thus a crucial task for stellar evolution theory, 
which requires to identify the detailed evolutionary paths 
of SNe Ia progenitors. 

Unlike core collapse supernovae, Type Ia supernovae (SNe Ia) are believed to occur
exclusively in binary systems (e.g. Livio~\cite{Livio01}). 
Although it is still unclear
which kinds of binary systems lead to SNe Ia, non-degenerate stars
such as main sequence stars, red giants or helium stars are often assumed
as white dwarf companion 
(e.g. Li \& van den Heuvel~\cite{Li97}; Hachisu et al.~\cite{Hachisu99}; Langer et al.~\cite{Langer00}; Yoon \& Langer~\cite{Yoon03}).
The white dwarf is then assumed to grow to the Chandrasekhar limit
by mass accretion from its companion, with accretion rates  
which allow steady shell hydrogen and 
helium burning (\Mdot{} $\gsim 10^{-7}$ \msyr{}).
An understanding of the physics of mass accretion is therefore
indispensable to investigate the evolution of accreting white dwarfs.

Although the mass accretion process in white dwarfs
has been discussed by many authors 
(e.g. Iben~\cite{Iben82}; Nomoto~\cite{Nomoto82}; Fujimoto \& Sugimoto~\cite{Fujimoto82}; 
Saio \& Nomoto~\cite{Saio85},~\cite{Saio98}; Kawai et al.~\cite{Kawai88}; 
Cassisi et al.~\cite{Cassisi98}; Piersanti et al.~\cite{Piersanti00};
Langer et al.~\cite{Langer02}), 
little attention was so far devoted to the effects of angular momentum accretion 
and the ensuing white dwarf rotation 
(see Sect.~\ref{sect:previous}).
As the evolution of stars can generally be affected by rotation
(e.g. Heger \& Langer~\cite{Heger00b}; Maeder \& Meynet~\cite{Maeder00}), 
this might be particularly so in accreting white dwarfs:
Since the transfered matter from the white dwarf companions 
may form a Keplerian disk which carries a large amount of angular momentum,
the resultant accretion of angular momentum will lead to the spin-up 
of the white dwarf (e.g. Durisen~\cite{Durisen77}; Ritter~\cite{Ritter85}; 
Narayan \& Popham~\cite{Narayan89}; 
Langer et al.~\cite{Langer00},~\cite{Langer02},~\cite{Langer03}). 
The observation that white dwarfs in cataclysmic variables rotate much 
faster than isolated ones (Sion~\cite{Sion99}) provides 
evidence for accreting white dwarfs indeed being spun up. 
Rapidly rotating progenitors may also lead to aspherical explosions,
which may give rise to the observed
polarization of SNe Ia (Howell et al.~\cite{Howell01}; Wang et al.~\cite{Wang03}).
 
Here, we make an effort to investigate in detail the possibility of
angular momentum accretion, and the role of the various 
rotationally induced hydrodynamic instabilities in transporting 
angular momentum into the white dwarf core, and in establishing
the the pre-explosion angular momentum profile.  
The remainder of this paper is organized as follows.
We evaluate the possible mechanisms for angular momentum 
transport in accreting white dwarfs in Sect.~\ref{sect:angmom}. 
Our approach to the problem, including the numerical methods and physical assumptions,
is discussed in Sect.~\ref{sect:simulation}, 
where previous papers on rotating white dwarf models are also 
reviewed (Sect.~\ref{sect:previous}).
Numerical results are presented in Sect.~\ref{sect:results}, 
with the emphasis on the process of angular momentum transport 
in the white dwarf interior.
Pre-explosion conditions of accreting white dwarfs and 
their implications for the diversity of SNe Ia are 
discussed in Sect.~\ref{sect:final}, ~\ref{sect:diversity} and
~\ref{sect:explosion}.
The possibility of detecting gravitational waves
from SNe Ia progenitors is examined in Sect.~\ref{sect:gwr}.
Our conclusions are summarized in Sect.~\ref{sect:conclusion}.

\section{Angular momentum transport in white dwarfs}\label{sect:angmom}

Inside a white dwarf, angular momentum can be transported by  
Eddington Sweet circulations
and by turbulent diffusion induced by hydrodynamic instabilities.
Our discussion below is limited to angular momentum transport
in the vertical direction.

\subsection{Eddington-Sweet circulation}\label{sect:ES}

Eddington-Sweet circulations are induced by the thermal 
imbalance between the equator and the poles of a rotating star. 
Its time scale  is roughly given by 
$\tau_{\rm ES} \simeq \tau_{\rm KH} /\chi^2$ (Maeder \& Meynet~\cite{Maeder00}), 
where $\tau_{\rm KH}$ is the Kelvin-Helmholtz 
time scale and $\chi$ is the angular velocity normalized to the Keplerian value, 
i.e., $\chi = \omega/\omega_{\rm Kep}$. 
In a white dwarf which accretes at rates \Mdot $ > 10^{-7}$ \msyr{}, 
the surface luminosity reaches $10^4$ \Lsun{} 
due to compressional heating and  nuclear burning. The quantity 
$\chi$ is close to 1 in the  outer envelope as we shall see in ~\ref{sect:limit}.
As a result, we expect an Eddington Sweet circulation time scale
in the non-degenerate envelope
which is shorter or comparable to the accretion time scale:
\begin{align}\label{eq1}
\tau_{\rm ES} & \approx  \frac{G\Delta M^2 _{\rm env}}{RL \chi^2} \nonumber \\ 
 & \approx
    1.3 \cdot 10^3 {\rm yr}
      \frac{ (\Delta M_{\rm env}/ 0.05 {\rm M_{\odot}})^2 }
           { (R/0.01 {\rm R_{\odot}}) (L/10^4 {\rm L_{\odot}}) (\chi/0.8)^2}
.
\end{align}

In the degenerate core, on the other hand, 
we have $\tau_{\rm ES}  \gg \tau_{\rm acc}$ 
since the Kelvin-Helmholtz time scale is 
typically much larger than $10^{7}$~yr. 
Furthermore, $\chi$ is usually significantly smaller than 1
in the inner core:
\begin{align}\label{eq2}
\tau_{\rm ES} & \approx  \frac{GM^2 _{\rm core}}{R_{\rm core}L_{\rm core} \chi^2} \nonumber \\ 
 & \approx
    7 \cdot 10^9 {\rm yr}
      \frac{ (M_{\rm core}/ {\rm M_{\odot}})^2 }
           { (R_{\rm core}/0.005{\rm R_{\odot}}) (L_{\rm core}/10 {\rm L_{\odot}}) (\chi/0.3)^2}
\end{align}
Consequently, Eddington-Sweet circulations are only important 
in the white dwarf envelope.

\subsection{Shear instability}\label{sect:shear}

The dynamical shear instability (DSI) occurs when 
the energy of the shear motion
dominates over the buoyancy potential. In a rotating flow, the
linear condition for instability is given by
\begin{equation}\label{eq3}
\frac{N^2}{\sigma^2} < R_{\rm i,c} \approx 1/4,  
\end{equation}
where $R_{\rm i,c}$ denotes the critical Richardson number which is about 0.25, 
 $N^2$  the Brunt-V$\ddot{\rm a}$is$\ddot{\rm a}$l$\ddot{\rm a}$ frequency, 
and $\sigma$ the shear factor (Zahn~\cite{Zahn74}): 
\begin{equation}\label{eq4}
N^2  =  N^2 _{\rm T} + N^2 _{\mu}, 
\end{equation}
\begin{equation}\label{eq5}
N^2 _{\rm T}  =  \frac{g \delta}{H_P} (\nabla_{\rm ad} - \nabla),  ~~~
N^2 _{\mu}   =  \frac{g\varphi\nabla_{\mu}}{H_{\rm P}}
\end{equation}
and
\begin{equation}\label{eq6}
\sigma = \frac{\partial \omega}{\partial \ln r}~. 
\end{equation}
Here, $\omega$ is the angular velocity and other symbols have their usual meaning. 
The equation of state of electron degenerate matter is given 
by $P=P(\rho, T, \mu_{\rm e}, \mu_{\rm I})$ and the factor $\varphi \nabla_{\mu}$ in 
Eq.~\ref{eq5}
needs to be replaced by $\varphi_{\rm e} \nabla_{\mu_{\rm e}} + \varphi_{\rm I}\nabla_{\mu_{\rm I}}$,
where $\varphi_{\rm e} = (\partial \ln \rho/\partial \ln \mu_{\rm e})_{P,T,\mu_{\rm I}}$
and $\varphi_{\rm I} = (\partial \ln \rho/\partial \ln \mu_{\rm I})_{P,T,\mu_{\rm e}}$.
In a typical CO white dwarf, 
$N^2 _{\mu}$ is generally smaller than $N^2 _{\rm T}$
and has minor effects on the angular momentum transport as discussed in Sect.~\ref{sect:mugrad}.

\begin{figure}
\resizebox{\hsize}{!}{\includegraphics{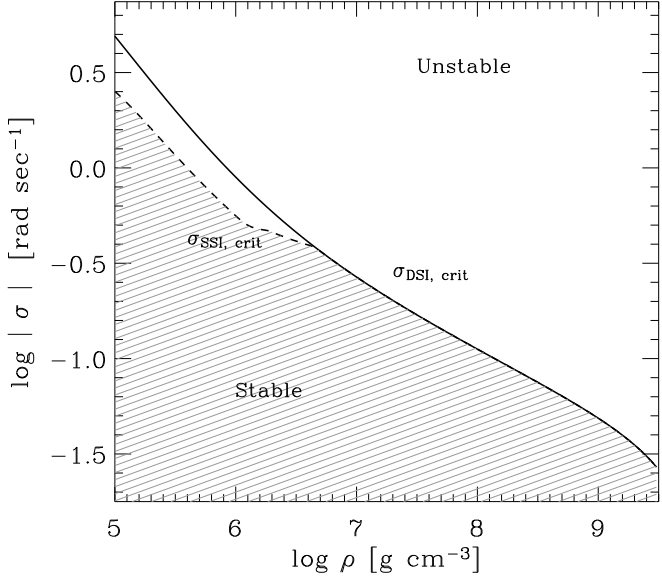}}
\caption{Threshold values of the shear factor $\sigma =(\partial \omega/\partial \ln r)$  
for the dynamical (solid line) and the secular (dashed line) shear instability.
Constant gravity and temperature, i.e., $g=10^9$ ${\rm cm/s^2}$, $T = 5\times10^7$ K, are assumed.
The chemical composition is also assumed to be constant, with $X_{\rm C}=0.43$
and $X_{\rm O}=0.54$. 
For the critical Richardson number, $R_{\rm i,c} = 1/4$ is employed. $R_{\rm e,c} = 2500$ is used
for the critical Reynolds number. 
}\label{fig:sigma}
\end{figure}

Under constant temperature and homogeneous chemical composition, which 
is a good approximation for the central region of a CO white dwarf, the
critical value of $\sigma$ for the DSI can be given as
\begin{align}\label{eq7}
& \sigma^2 _{\rm DSI, crit}  
 =  N^2/R_{\rm i, c}   \nonumber \\
& \simeq  0.2 \left(\frac{g}{10^9 {\rm cm/s^2}}\right) \left(\frac{\delta}{0.01}\right)
\left(\frac{H_{\rm P}}{8 \cdot 10^7 {\rm cm}} \right)^{-1} \left(\frac{\nabla_{\rm ad}}{0.4}\right) 
\end{align}
where $R_{\rm i, c}=0.25$ is assumed.
Fig.~\ref{fig:sigma} shows $|\sigma_{\rm DSI, crit}|$
as function of density, with other physical assumptions
as indicated in the figure caption. 
This figure shows that the condition for the DSI becomes more relaxed with increasing density, 
which is due to the buoyancy force becoming weaker with stronger degeneracy.
This implies that the highly degenerate core in a white dwarf is more susceptible 
to the dynamical shear instability than the outer envelope. 

The stability criterion for the shear instability 
can be relaxed if thermal diffusion reduces the buoyancy force such that
\begin{equation}\label{eq8}
 R_{\rm is,1} :=  \frac{\nu R_{\rm e,c}N_{\rm T}^2}{K\sigma^2} < R_{\rm i,c},
\end{equation}
where $R_{\rm e,c}$ is the critical Reynolds number,
$\nu$ is the kinetic viscosity and $K =(4acT^3)/(3C_{\rm p}\kappa\rho^2)$  is the
thermal diffusivity (e.g. Maeder \& Meynet~\cite{Maeder00}).
Since the $\mu$ gradient is not affected by the thermal diffusion, 
this condition needs to be supplemented 
by the following additional condition (Endal \& Sofia~\cite{Endal76}):
\begin{equation}\label{eq9}
R_{\rm is, 2} := \frac{N^2 _{\mu}}{\sigma^2} < R_{\rm i,c}
\end{equation}
The thermally induced shear instability
is often called 'secular shear instability' (SSI). 

\begin{figure}
\resizebox{\hsize}{!}{\includegraphics{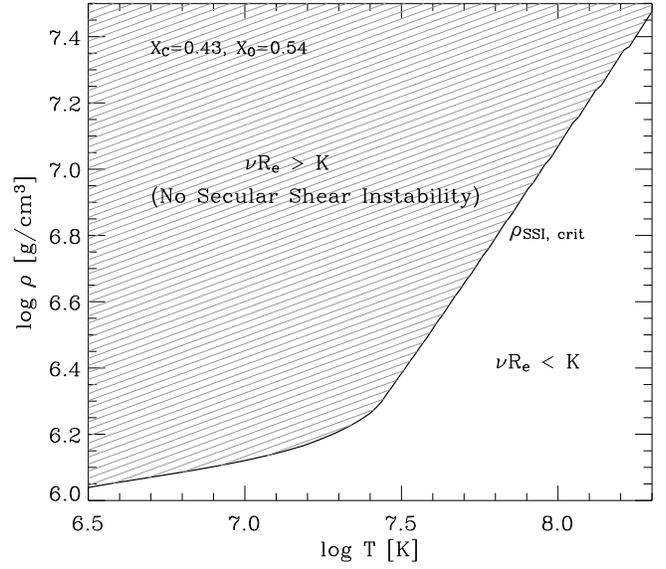}}
\caption{
The critical density $\rho_{\rm SSI, crit}$,
above which the secular shear instability can not occur, as 
function of temperature (solid line).
The region where the thermal diffusion time scale exceeds
the turbulent viscous time scale is
hatched. Abundances of $X_{\rm C}=0.43$ and $X_{\rm O}=0.54$
are assumed. $R_{\rm e, c} =2500$ is
used for the critical Reynolds number.
}\label{fig:rho_ssi}
\end{figure}

This instability can occur only when the thermal diffusion 
time scale is shorter than the turbulent viscous time scale 
(i.e., $\nu R_{\rm e,c} < K$), which is often the case
in non-degenerate stars. However, 
this condition is not always fulfilled in white dwarfs.
In Fig.~\ref{fig:rho_ssi}, the critical density  $\rho_{\rm SSI, crit}$
above which the secular
shear instability is not allowed is plotted as function of temperature. 
In this calculation, the ion and electron viscosity as well as
the radiative and conductive opacities are estimated
as described in Sect.~\ref{sect:method}. 
This figure indicates that the SSI may play a role only relatively
weakly degenerate regions.

The threshold value of the shear factor for this instability
is given by
$ \sigma^2 _{\rm SSI, crit}  
 =  \nu R_{\rm e,c} N^2 _{\rm T}/(KR_{\rm i, c}) $
if \nvmu{} = 0,
and plotted
in Fig~\ref{fig:sigma}, 
with the assumptions as indicated in the figure caption. 
As shown in the figure and as discussed above, 
thermal diffusion can not weaken the restoring buoyancy force for $\rho 
\gsim 5 \times 10^6$ ${\rm g/cm^3}$ under the chosen conditions,
and only the dynamical shear instability can occur for higher densities.

Zahn (\cite{Zahn92}) derives the diffusion coefficient for the SSI as
\begin{equation}\label{eq10}
D_{\rm SSI} = \frac{1}{3}\frac{K\sigma^2R_{\rm i,c}}{N^2}.
\end{equation}
The time scale of the SSI
when $\sigma=\sigma_{\rm DSI, crit}$
thus becomes simply 
the thermal diffusion time scale:
$\tau_{\rm SSI}~({\rm at}~\sigma = \sigma_{\rm DSI,crit}) \approx
3R^2/K$.  
In the degenerate core of a CO white dwarf 
with $\rho \lsim \rho_{\rm SSI,crit}$, 
we have $\tau_{\rm SSI} \approx 7.6 \times 10^6 
(R_{\rm core}/2\cdot10^8 ~{\rm cm})^2/(K/500 ~{\rm cm^2s^{-1}})$ yr.

As consequence of these considerations,
any shear motion with $\sigma > \sigma_{\rm DSI, crit}$ will decay
such that $\sigma$ approaches $\sigma_{\rm DSI, crit}$ on the dynamical 
time scale. Further angular momentum transport will operate
on the thermal diffusion time scale until $\sigma$ reaches $\sigma_{\rm SSI, crit}$ 
for $ \rho \lsim \rho_{\rm SSI, crit}$.
For $\sigma < \sigma_{\rm SSI, crit}$ and $\sigma<\sigma_{\rm DSI,crit}$
(hatched region in Fig.~\ref{fig:sigma}), 
angular momentum will be transported only
via the electron viscosity or via Eddington Sweet circulations on much
longer time scales
(unless other kinds of instabilities are invoked).
Therefore, if we consider accretion at rates required by the single degenerate
SN Ia progenitor scenario (\Mdot$ > 10^{-7}$ \msyr), 
the degree of shear may not be far from 
the threshold value for the dynamical shear instability throughout the 
degenerate white dwarf interior. 
This conjecture is confirmed by the numerical results 
presented in Sect.~\ref{sect:spin}.

\subsection{GSF and magnetic instabilities}\label{sect:GSF}

The Goldreich, Schubert and Fricke instability (GSF instability) 
can be induced if a star is in a baroclinic condition 
(Goldreich \& Schubert~\cite{Goldreich67}; Fricke~\cite{Fricke68}). 
In an accreting white dwarf,
this instability may be important in the non-degenerate envelope, but is likely
suppressed in the degenerate core where the barotropic condition
will be retained through a dynamical meridional circulation 
(Kippenhahn \& M\"ollenhoff~\cite{Kippenhahn74}; see also Durisen~\cite{Durisen77}). 
Magnetic instabilities such as the Tayler instability (Spruit~\cite{Spruit02}) 
may be potentially important, and their role will be investigated in the near future. 
In the present study, we restrict our discussions to non-magnetized white dwarfs.

\section{Simulations of accreting white dwarfs}\label{sect:simulation}

For the detailed investigation of the angular momentum transport 
and its role in the evolution of accreting white dwarfs, 
we have simulated accreting CO white dwarfs with various
initial masses and accretion rates.  

\subsection{Previous work on rotating white dwarfs}\label{sect:previous}

In order to compare our results with previously obtained ones, we briefly discuss the
models of rotating white dwarfs in the literature here.
Most rotating white dwarf equilibrium models have been constructed 
assuming a barotropic equation of state and axial symmetry
for both, rigidly  
and differentially  rotating cases
(e.g. James~\cite{James64}; Roxburgh~\cite{Roxburgh65}; Monaghan~\cite{Monaghan66}; 
Ostriker~\cite{Ostriker66}; Anand~\cite{Anand68}; Ostriker \& Mark~\cite{Ostriker68b}; 
Hachisu~\cite{Hachisu86}).   
These models have mainly been used to investigate the stability of 
rapidly rotating
zero temperature white dwarfs (e.g. James~\cite{James64}; 
Lynden-Bell \& Ostriker~\cite{Lynden67}; 
Ostriker \& Bodenheimer~\cite{Ostriker68a}; Durisen~\cite{Durisen75b}; 
Durisen \& Imammura~\cite{Durisen81}).
One of the main results of these studies is that the Chandrasekhar 
limit increases slightly in rigidly rotating white dwarfs ($\sim$ 1.48 \Msun), while
differentially rotating white dwarfs can be dynamically stable even
with masses up to $\sim$ 4.0 \Msun{}. 
It was also found that differentially rotating white dwarfs
become secularly unstable to gravitational wave radiation
if the ratio of the rotational energy to the gravitational energy 
exceeds a certain limit 
(see Sect.~\ref{sect:final} for more detailed discussions).

The evolution of rotating white dwarfs was investigated
by Durisen (\cite{Durisen73a},~\cite{Durisen73b},~\cite{Durisen75a}) for the first time. 
Evolution and spin-up of mass accreting white dwarfs were probed
by Durisen (\cite{Durisen77}).
In his studies based on two-dimensional barotropic differentially rotating
white dwarf models, Durisen assumed that angular momentum in the white dwarf
interior is only transported by electron viscosity, and angular
momentum transport time scales of the order of $10^{10}\,$yr have been
obtained. 
In all these models, 
the thermal history and the energy transport in the white dwarf matter
were neglected, assuming zero temperature.

Recently, Piersanti et al. (\cite{Piersanti03}) investigated the effect of rotation on the 
thermal evolution of CO-accreting white dwarfs,
using a one dimensional stellar evolution code. 
They also discussed angular momentum loss by gravitational 
wave radiation (cf. Sect.~\ref{sect:final}).
However, the detailed history of the angular momentum 
transport inside the white dwarf was neglected, 
but rigid body rotation was assumed.  
I.e., while Durisen assumed much slower angular momentum redistribution
by restricting his considerations to electron diffusion, 
Piersanti et al. assume instant angular momentum redistribution with
a maximum efficiency.

Uenishi et al. (\cite{Uenishi03}) did consider a finite spin-up time of
accreting white dwarfs by dividing their two-dimensional
white dwarf models into a non-rotating central core and a 
fast rotating outer envelope. However,
rigid body rotation is again assumed for the spun-up region, 
introducing a discontinuity in the angular velocity profile.

In the present study on the evolution of accreting white dwarfs, 
we consider 
the effects of the accretion induced heating and energy transport, 
the angular momentum transport by various instabilities,
and the effect of rotation on the white dwarf structure.
However, our numerical models are calculated 
in a one dimensional approximation as described in the next section, 
and thus the structure of rapidly rotating white dwarfs 
is described less accurate than in multi-dimensional models mentioned above. 
Furthermore, by assuming CO-accretion we do not consider 
effects of nuclear shell burning. These, and the feedback between
rotation and nuclear shell burning, as well as rotationally induced
chemical mixing, are discussed in a separate paper 
(Yoon et al.~\cite{Yoon04c}).  The effects of 
rotation with respect to the off-center helium detonation models 
 -- i.e., helium accreting CO white dwarf models with \Mdot{} $= \sim 10^{-8}$~\msyr{} -- 
are investigated in Yoon \& Langer (\cite{Yoon04b}).

\subsection{Numerical method}\label{sect:method} 

We use a hydrodynamic stellar evolution code (Langer et al.~\cite{Langer88}) for 
the accreting white dwarf model calculations. 
Radiative opacities are taken from Iglesias \& Rogers (\cite{Iglesias96}).
For the electron conductive opacity, 
we follow  Hubbard \& Lampe (\cite{Hubbard69}) 
in the non-relativistic  case, and Canuto (\cite{Canuto70}) in the relativistic case. 
The accreted matter is assumed to have the same entropy as
that of the surface of the accreting white dwarf, and the accretion
induced compressional heating is treated as in Neo et al (\cite{Neo77}).

The effect of rotation on the stellar structure is approximated
in one dimension
by introducing the factors $f_{\rm P}$ and $f_{\rm T}$ 
in the momentum and energy conservation equations
(cf. Heger et al.~\cite{Heger00a}), 
following the method of Kippenhahn \& Thomas (\cite{Kippenhahn70}) 
and Endal \& Sofia (\cite{Endal76}).
This method is well suited for the case 
where the effective gravity can be derived from an effective potential.
This is indeed a good  assumption
for a CO white dwarf, since it should be in a barotropic condition
except for the non-degenerate outer envelope. 
Note that this method is also applicable in the case
of shellular rotation as discussed by Meynet \& Maeder (1997). 

However, our method of computing
the effective gravitational potential in a rotating star limits
the accuracy of our results for very rapid rotation. 
The potential is expanded in terms of spherical harmonics,
of which we only consider terms up to the second order
(cf., Kippenhahn \& Thomas~\cite{Kippenhahn70}).
Fliegner (\cite{Fliegner93}) showed this method
to accurately reproduce the shapes of rigidly rotating polytropes
up to a rotation rate of about 60\% critical, corresponding
to correction factors of $f_{\rm P}\simeq 0.75$ and $f_{\rm T} \simeq 0.95$
in the stellar structure equations (cf. Heger et al.~\cite{Heger00a}).
We therefore limit these factors to the quoted values,
with the consequence that we underestimate the effect
of the centrifugal force in layers which rotate faster than
about 60\% critical. 
As in our models the layers near the surface 
are always close to critical rotation,
the stellar radius of our models may be underestimated.
Furthermore, our models are per definition rotationally symmetric.
Therefore, we are in principle 
unable to investigate the onset of triaxial instabilities, which
may affect the final fate of a rapidly rotating white dwarf (Sect.~\ref{sect:final}).

The angular momentum transport induced by the instabilities 
mentioned in Sect.~\ref{sect:angmom} is described as a diffusion 
process (Heger et al.~\cite{Heger00a}), while this approximation neglects
the possibility of advective angular momentum redistribution by Eddingtion Sweet circulations  
(Maeder \& Meynet~\cite{Maeder00}).
Note also that this approach is based on the assumption of shellular 
rotation (see Heger et al.~\cite{Heger00a} for a detailed discussion), which
might not be appropriate for white dwarfs. 
Nevertheless, our models can represent the case of cylindrical 
rotation to some degree, since most of the total angular momentum
is confined to layers near the equatorial plane in both cases. 
Since the dynamical shear instability is important in the present study, 
the diffusion solver has been improved as a non-linear process as explained in Yoon 
(\cite{Yoon04e}), 
in order to properly deal with such a fast angular momentum redistribution 
as on a dynamical time scale during the secular evolution of 
accreting white dwarfs.
Diffusion coefficients for each of the instabilities
are taken from Heger et al. (\cite{Heger00a}) and Heger and Langer (\cite{Heger00b}), with some
modifications as follows. 

As mentioned in Sect.~\ref{sect:GSF}, the GSF instability may be suppressed
in the degenerate barotropic core, and we describe this effect as 
\begin{equation}\label{eq11}
D_{\rm GSF}=D_{\rm GSF}^*\left(1-{\rm min}\left[1,\frac{P_{\rm e}}{P_{\rm ideal}}\right]\right)~.
\end{equation}
Here, $D_{\rm GSF}^*$ is the diffusion coefficient for the GSF instability
as estimated in Heger et al. (\cite{Heger00a}). $P_{\rm e}$ is the electron pressure
for complete non-relativistic degeneracy, and $P_{\rm ideal}$ is the ideal
gas pressure:
\begin{equation}\label{eq12}
P_{\rm e} = 9.91\cdot10^{12}\left(\frac{\rho}{\mu_{\rm e}}\right)^{5/3}, 
~~~ P_{\rm ideal} = \frac{\rho kT}{\mu m_{\rm H}}. 
\end{equation}
For the secular instability, Zahn's (\cite{Zahn92}) prescription is employed (Eq.~\ref{eq10}), 
with a correction factor as in Heger et al. (\cite{Heger00a}) 
to ensure a smooth turn-on of the instability and to relate its strength 
to the deviation from the Richardson criterion:
\begin{equation}\label{eq13}
D_{\rm SSI} = \frac{1}{3}\frac{K\sigma^2R_{\rm i,c}}{N^2} 
\left(1 - \frac{{\rm max}\{R_{\rm is,1},~R_{\rm is,2}\}}{R_{\rm i,c}}\right)^2 ~,
\end{equation}
where $R_{\rm is,1}$ and $R_{\rm is,2}$ are defined in Eq.~\ref{eq8} and~\ref{eq9}. 
As already discussed in Sect.~\ref{sect:shear}, 
the viscosity is an important factor for the stability of the secular shear
instability.
The former way of computing the ion viscosity in Heger et al. (\cite{Heger00a}) 
and Heger \& Langer (\cite{Heger00b}), 
following Spitzer (\cite{Spitzer62}),
is not well adapted for matter in the ion liquid state: here we follow
the work of Wallenborn and Bauss (\cite{Wallenborn78}), 
as described in Itoh et al. (\cite{Itoh87}):
\begin{equation}\label{eq14}
\eta_{\rm i} = 5.53\cdot10^3 Z A^{-1/3}(\rho/10^6)^{5/6}\eta^*~~{\rm (g~cm^{-1}~s^{-1})} ~.
\end{equation}
We constructed a fitting formula for the dimensionless parameter $\eta^*$ from the table in Itoh et al. as
\begin{align}\label{eq15}
\eta^* = & 
 -0.016321227+ 1.0198850~\Gamma^{-1.9217970} \nonumber \\
 & +0.024113535~\Gamma^{0.49999098} ~~,
\end{align}
where $\Gamma = (Ze)^2/[(3/4\pi n_{\rm i})^{1/3} kT]$.
The electron viscosity, which is dominant over the ion viscosity 
in degenerate matter, is estimated according to
Nandkumar \& Pethick (\cite{Nandkumar84}). 

When differential rotation is present in a star, 
rotational energy is dissipated through frictional heating. 
We estimate the rotational energy dissipation rate as 
(Kippenhahn \& Thomas~\cite{Kippenhahn78}; Mochkovitch \& Livio~\cite{Mochkovitch89})  
\begin{equation}\label{eq16}
\epsilon_{{\rm diss}} = \frac{1}{2}\nu_{\rm turb} \sigma^2  ~~~{\rm (erg~g^{-1}~sec^{-1})} ~,
\end{equation}
where $\nu_{\rm turb}$ is the turbulent viscosity, which
is the sum of all the diffusion coefficients related to the rotationally
induced instabilities considered in the present study (Heger et al.~\cite{Heger00a}). 

As discussed in Sect.~\ref{sect:shear},
the calculation of $\varphi_{\mu_{\rm e}}$ and $\varphi_{\mu_{\rm I}}$
is required for 
considering chemical composition effects in degenerate matter 
on the rotationally induced instabilities.
We can make use of the thermodynamic relation:
\begin{equation}\label{eq17}
\varphi_{\rm e}\nabla_{\mu_{\rm e}}+\varphi_{\rm I}\nabla_{\mu_{\rm I}} 
 = \frac{d\ln \rho}{d\ln P} - \alpha + \delta \nabla ~, 
\end{equation}
where all the thermodynamic quantities on the right hand side 
are obtained from the equation of state used in our
stellar evolution code (Blinnikov et al.~\cite{Blinnikov96}).

\subsection{Physical assumptions}\label{sect:assumption} 

\begin{table}[t]
\begin{center}
\caption{Physical quantities of the initial white dwarf models:
 mass, surface luminosity, central temperature, central density, 
radius and rotation velocity}\label{tab:init}
\begin{tabular}{c c c c c c }
\hline \hline
$M_{\rm WD, init}$  &  $L_{\rm s, init}$ & $T_{\rm c, init}$ & $\rho_{\rm c, init}$ & $R_{\rm WD, init}$ & $v_{\rm rot, init} $ \\
    \Msun           &        \Lsun       &     $10^8$ K      & $10^7~{\rm g/cm^3}$  &  \Rsun             &     km/s \\
\hline
    0.8             &   0.80             &     0.79          &  0.97                 & 0.0115            &    10 \\
    0.9             &   0.88             &     0.85          &  1.68                 & 0.0100            &    10 \\
    1.0             &   0.81             &     0.94          &  3.14                 & 0.0086            &    10 \\
\hline
\end{tabular}
\end{center}
\end{table}

\begin{figure}
\resizebox{\hsize}{!}{\includegraphics{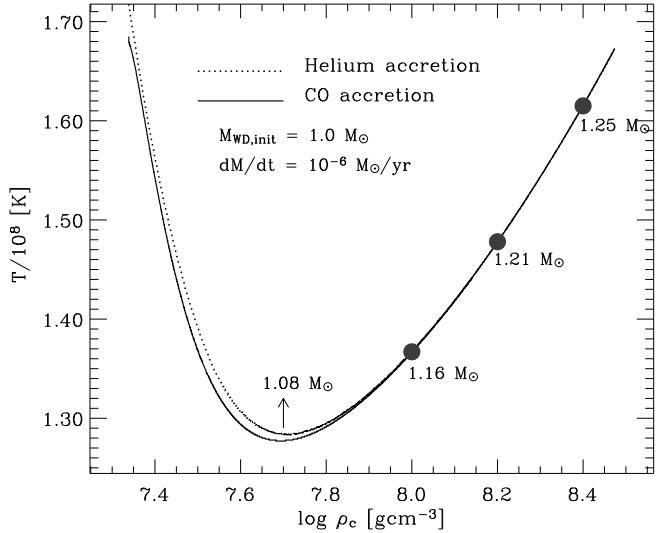}}
\caption{
Evolution of central density and temperature
in non-rotating accreting CO white dwarfs. The initial mass
is 1.0 \Msun{} and the accretion rate is \mdotd{} \msyr.
The dotted line denotes the case of helium accretion, while
the dashed line the case of carbon-oxygen accretion. 
The initial central temperature was $T_{\rm c, init} = 2.29 \cdot 10^8$ K
for the helium accreting white dwarf and 
and $T_{\rm c, init} = 1.69 \cdot 10^8$ K  for the CO accreting one.
The numbers next to the filled circles denote the
white dwarf mass in solar masses.
}\label{fig:dem}
\end{figure}

Our CO white dwarf models start mass accretion at 
three different initial masses: 0.8, 0.9 and 1.0 \Msun. 
Isolated white dwarfs are observationally found to rotate slowly
($\mathnormal{v}_{\rm WD} \lsim 40 $ km/s ; Heber et al.~\cite{Heber97}; 
Koester et al.~\cite{Koester98}; Kawaler~\cite{Kawaler03}), as also
predicted by stellar evolution models (Langer et al.~\cite{Langer99}).
We thus consider slow rigid rotation in our initial models, 
and the surface velocity at the white dwarf equator is set to 10 km/s. 
Other physical quantities of the initial models are 
summarized in Table~\ref{tab:init}.

As for the accretion rate, 
four different values are considered: \mdota, \mdotb, \mdotc{}
and \mdotd{} \msyr. These accretion rates are chosen in the context 
of the single degenerate Chandrasekhar mass scenario for SNe Ia progenitors,
in which steady nuclear shell burning of hydrogen or helium is assumed 
(Nomoto~\cite{Nomoto82}; Li \& van den Heuvel~\cite{Li97}; Langer et al.~\cite{Langer00}; 
Yoon \& Langer~\cite{Yoon03}). 
The accreted matter is carbon-oxygen enriched such that
$X_{\rm C}=X_{\rm O} = 0.487$. 
In this way, we assume that the accreted
hydrogen or helium is immediately converted to carbon and oxygen by
shell burning. 
This assumption, however, does not affect 
the advanced thermal evolution of 
the degenerate core of a white dwarf as demonstrated in Fig.~\ref{fig:dem}.
In this figure, the evolution of the central density and temperature
of non-rotating white dwarf which accrete matter at a rate
of with \Mdot = \mdotd{} \msyr{}
is shown for two cases.
One case assumes CO accretion, the other case assumes helium accretion.
In the latter case,
we have followed the helium
shell burning and thereby induced thermal pulses up to \Mwd = 1.2 \Msun.
Although a slight difference appears  
when \Mwd $\lsim$ 1.1 \Msun{}, which is due to the different initial temperatures
as indicated in the figure caption, 
the two sequences converge when \Mwd $\gsim$ 1.1 \Msun{} 
due to the dominance of the compressional heating for the thermal
structure. Therefore, we can 
assume safely that the thermal evolution of the white dwarf interior 
does not change significantly even if we neglect
the shell burning, as long as rapid accretion is ensured.

Since non-magnetic white dwarfs in close binary systems 
are thought to accrete matter through a Keplerian disk,
the accreted matter may carry angular momentum 
of about the local Keplerian value at the white dwarf equator
(cf. Durisen~\cite{Durisen77}; Ritter~\cite{Ritter85}). 
Langer et al. (\cite{Langer00}) pointed out that a white dwarf
will reach overcritical rotation well before
growing to the Chandrasekhar limit 
if the white dwarf gains angular momentum with
the Keplerian value,
when the assumption of rigid body rotation is made
(see also Livio \& Pringle~\cite{Livio98}; Uenishi et al.~\cite{Uenishi03}). 
Our preliminary calculations show that
overcritical rotation may be reached even earlier at 
the white dwarf surface, 
as the angular momentum transport time scale is
finite (Yoon \& Langer~\cite{Yoon02}).
Interestingly, 
Paczy\'nski (\cite{Paczynski91}) and Popham \& Narayan (\cite{Popham91})
found a possibility for the angular momentum 
to be transported from the white dwarf into the disk
by viscous effects when the white dwarf rotates
at near break-up velocity, without preventing the continued
efficient mass accretion.
Based on this picture, we limit the angular momentum 
gain such that no angular momentum is allowed to be transfered
into an accreting white dwarf when its surface rotates
at the Keplerian value at its equator:
\begin{equation}\label{eq18}
j_{\rm acc} = \left\{ \begin{array}{ll}
f \cdot j_{\rm K} & \textrm{if $ v_{\rm{s}} < v_{\rm{K}}$ } \\
0 &  \textrm{if $ v_{\rm{s}} = v_{\rm{K}}$ } \end{array} \right.
\end{equation}
where $j_{\rm acc}$ denotes the specific angular momentum carried by the accreted matter,
$v_{\rm s}$ the surface velocity of the accreting white dwarf at its equator,
$v_{\rm K}$ and $j_{\rm K}$ the Keplerian value of the rotation velocity
and the specific angular momentum at the white dwarf equator, respectively:
$v_{\rm K} = (GM_{\rm WD}/R_{\rm WD})^{1/2}$ and $j_{\rm K} = v_{\rm K}R_{\rm WD}$. 
The dimensionless parameter $f$ determines which fraction of the Keplerian value
is actually accreted in a white dwarf when $v_{\rm s} < v_{\rm K}$. 

As mentioned earlier, we do not consider nuclear shell burning
in our simulations. Although this may not change the thermal
history in the white dwarf core, the surface conditions can be
affected by this simplification: the white dwarf envelope may be hotter and more extended
with shell burning, which may lead to a stronger Eddington Sweet circulation and
different values for $j_{\rm K}$. 
In order to investigate how the results are affected by this uncertainty, 
we consider three different values for $f$: 0.3, 0.5 and 1.0.

In calibrating the chemical mixing efficiencies in massive stars,
Heger et al. (\cite{Heger00a}) introduced the factor $f_{\mu}$ such that
\nabmu{} is replaced by $f_{\mu}\nabla_{\mu}$, with
$f_{\mu} = 0.05$ giving the best reproduction of observed enhancements of nitrogen
at the stellar surface in the mass range 10 \Msun{} to 20 \Msun. 
Although this result is deduced from massive stars and might not apply
to white dwarfs, we keep the value of $f_{\mu} = 0.05$ for most
of our models. However, in order to check the sensitivity 
of our results to this parameter,  
models with $f_{\mu} = 1.0$ are also calculated (Sect.~\ref{sect:mugrad}).

Table~\ref{tab:result1} gives an overview of the computed model
sequences.
The first column gives the model sequence designation, 
where the letters A, B and C
denote the cases with $f=$ 1.0, 0.5 and 0.3 respectively, all
for a fixed $f_{\mu}$ of 0.05.
Model sequences with the index D have $f=1.0$ and $f_{\mu}=1.0$. 
The index T indicates test sequences where
the effects of rotational energy dissipation (Eq.~\ref{eq16}) are neglected.

% -------------------------------------------------------------------------------------------------------------
% -------------------------------------------------------------------------------------------------------------
% -------------------------------------------------------------------------------------------------------------
% -------------------------------------------------------------------------------------------------------------
% -------------------------------------------------------------------------------------------------------------

\begin{table*}[!]
\begin{center}
\caption{ Properties of the computed model sequences. 
The first column gives the system number. 
The other columns have the following meanings.
$M_{\rm init}$: initial mass, \Mdot: accretion rate,  
$f$: fraction of $j_{\rm K}$ of the accreted matter (see Eq.~\ref{eq18}),
$f_{\mu}$ : efficiency parameter of the $\mu$ gradient,
\Erot : rotational energy dissipation by friction, 
$M_{0.10}$: white dwarf mass when the ratio of the rotational energy to the gravitational
potential energy ($T/W$) reaches 0.10. The rest physical quantities are also  
estimated at this point, i.e., when $T/W = 0.10$.
$\rho_{\rm c, 0.10} $ and $T_{\rm c, 0.10}$ :
central density and temperature. $R_{0.10}$: radius of white dwarf 
defined on the sphere of the volume of the equipotential surface. 
$J_{0.10}$ : total angular momentum,  $\Omega_{0.10}$: moment of inertia weighted mean of 
angular velocity. 
}\label{tab:result1}
\begin{tabular}{r | c c c  c c | c c c c c c }
\hline \hline
No.& $M_{\rm init}$ & \Mdot & $f$ & $f_{\mu}$ & $\dot{E}_{\rm rot}$ &
$M_{0.10}$ & $\rho_{\rm c, 0.10} $ & $T_{\rm c, 0.10}$ & $R_{0.10}$ & $J_{0.10}$ & $\Omega_{0.10}$ \\ 
\hline
    &        & $10^{-7}$ &  & & &   &  $10^8$ & $10^8$ & 0.01&  $10^{50}$ &         \\
    & \Msun  &   \msyr   &  & & &  \Msun &   ${\rm g/cm^3}$ &  K  & \Rsun  & erg s & rad/s    \\
\hline
A1 & 0.8 & 3.0   & 1.0 & 0.05 & Yes &  1.16 & 0.29 & 0.92 & 1.22 & 1.10 & 0.68  \\
A2 & 0.8 & 5.0   & 1.0 & 0.05 & Yes &  1.18 & 0.32 & 0.97 & 1.26 & 1.13 & 0.70   \\          
A3 & 0.8 & 7.0   & 1.0 & 0.05 & Yes & 1.19 & 0.33 & 1.01 & 1.34 & 1.14 & 0.69   \\          
A4 & 0.8 & 10.0  & 1.0 & 0.05 & Yes & 1.21 & 0.35 & 1.07 & 1.39 & 1.17 & 0.71    \\          
A5 & 0.9 & 3.0   & 1.0 & 0.05 & Yes & 1.28 & 0.58 & 1.02 & 1.02 & 1.18 & 0.95  \\          
A6 & 0.9 & 5.0   & 1.0 & 0.05 & Yes & 1.30 & 0.64 & 1.11 & 1.03 & 1.21 & 0.99   \\          
A7 & 0.9 & 7.0   & 1.0 & 0.05 & Yes &1.32 & 0.69 & 1.17 & 1.03 & 1.23 & 1.02   \\
A8 & 0.9 & 10.0  & 1.0 & 0.05 & Yes & 1.34 & 0.75 & 1.23 & 1.05 & 1.25 & 1.06  \\
A9 & 1.0 & 3.0   & 1.0 & 0.05 & Yes &1.41 & 1.30 & 1.14 & 0.82 & 1.26 & 1.40  \\
A10& 1.0 & 5.0   & 1.0 & 0.05 & Yes &1.42 & 1.39 & 1.22 & 0.83 & 1.25 & 1.43  \\
A11& 1.0 & 7.0   & 1.0 & 0.05 & Yes & 1.43 & 1.52 & 1.28 & 0.82 & 1.27 & 1.49  \\
A12& 1.0 & 10.0  & 1.0 & 0.05 & Yes &1.45 & 1.67 & 1.36 & 0.83 & 1.28 & 1.56  \\
\hline
B1  & 0.8 & 3.0   & 0.5 & 0.05 & Yes & 1.21 & 0.36 & 0.96 & 1.17 & 1.17 & 0.76  \\
B2  & 0.8 & 5.0   & 0.5 & 0.05 & Yes & 1.25 & 0.42 & 1.04 & 1.20 & 1.22 & 0.81  \\
B3  & 0.8 & 7.0   & 0.5 & 0.05 & Yes & 1.26 & 0.44 & 1.09 & 1.24 & 1.23 & 0.82  \\
B4  & 0.8 & 10.0  & 0.5 & 0.05 & Yes &  1.30 & 0.52 & 1.16 & 1.23 & 1.27 & 0.89 \\
B5  & 0.9 & 3.0   & 0.5 & 0.05 & Yes &1.32 & 0.69 & 1.03 & 0.97 & 1.24 & 1.04   \\
B6  & 0.9 & 5.0   & 0.5 & 0.05 & Yes & 1.35 & 0.79 & 1.12 & 0.97 & 1.27 & 1.11  \\
B7  & 0.9 & 7.0   & 0.5 & 0.05 & Yes & 1.38 & 0.90 & 1.12 & 0.96 & 1.29 & 1.18  \\
B8  & 0.9 & 10.0  & 0.5 & 0.05 & Yes &  1.41 & 1.04 & 1.27 & 0.97 & 1.32 & 1.26 \\
B9  & 1.0 & 3.0   & 0.5 & 0.05 & Yes & 1.45 & 1.61 & 1.18 & 0.78 & 1.30 & 1.56  \\
B10 & 1.0 & 5.0   & 0.5 & 0.05 & Yes & 1.46 & 1.76 & 1.27 & 0.78 & 1.30 & 1.62  \\
B11 & 1.0 & 7.0   & 0.5 & 0.05 & Yes & 1.48 & 2.00 & 1.34 & 0.78 & 1.31 & 1.72  \\
B12 & 1.0 & 10.0  & 0.5 & 0.05 & Yes &1.51 & 2.33 & 1.44 & 0.77 & 1.32 & 1.84  \\
\hline
C1 & 0.8 & 3.0 & 0.3 & 0.05 & Yes & 1.53 & 1.55 & 1.17 & 0.77 & 1.48 & 1.57 \\  
C2 & 0.8 & 5.0 & 0.3 & 0.05 & Yes & 1.50 & 1.25 & 1.19 & 0.84 & 1.46 & 1.41 \\  
C3 & 0.8 & 7.0 & 0.3 & 0.05 & Yes & 1.46 & 1.05 & 1.20 & 0.90 & 1.44 & 1.29 \\
C4 & 0.8 & 10.0& 0.3 & 0.05 & Yes & 1.42 & 0.85 & 1.21 & 1.01 & 1.40 & 1.15 \\
C9 & 1.0 & 3.0 & 0.3 & 0.05 & Yes & 1.73 & 19.0 & 2.16 & 0.42 & 1.25 & 5.03 \\
C10& 1.0 & 5.0 & 0.3 & 0.05 & Yes & 1.72 & 16.2 & 2.04 & 0.44 & 1.27 & 4.65 \\
C11& 1.0 & 7.0 & 0.3 & 0.05 & Yes & 1.72 & 14.3 & 2.06 & 0.46 & 1.29 & 4.38 \\
C12& 1.0 & 7.0 & 0.3 & 0.05 & Yes & 1.71 & 12.1 & 2.01 & 0.49 & 1.31 & 4.04 \\
\hline
D2  & 0.8 & 5.0   & 1.0 & 1.00 & Yes & 1.18 & 0.32 & 0.97 & 1.27 & 1.13 & 0.69  \\
D6  & 0.9 & 5.0   & 1.0 & 1.00 & Yes & 1.30 & 0.64 & 1.11 & 1.03 & 1.21 & 0.99  \\
\hline
T2  & 0.8 & 5.0   & 1.0 & 0.05 & No  & 1.12 & 0.27 & 0.95 & 1.24 & 1.02 & 0.63  \\
T6  & 0.9 & 5.0   & 1.0 & 0.05 & No  & 1.26 & 0.56 & 1.03 & 0.97 & 1.15 & 0.93  \\
T10 & 1.0 & 5.0   & 1.0 & 0.05 & No  & 1.40 & 1.24 & 1.14 & 0.81 & 1.24 & 1.37  \\
\hline
\hline
\end{tabular}
\end{center}
\end{table*}

% -------------------------------------------------------------------------------------------------------------
% -------------------------------------------------------------------------------------------------------------
\begin{table*}[!]
\begin{center}
\caption{
Continued from Table~\ref{tab:result1}. The symbols in the columns have the same
meaning with those in Table~\ref{tab:result2}, but when $T/W = 0.14$ and $T/W = 0.18$ 
for the indices 0.14 and 0.18, respectively. 
}\label{tab:result2}
\begin{tabular}{r | c c c c c c  | c c c c c c  }
\hline \hline
No.& 
$M_{0.14}$ & $\rho_{\rm c, 0.14} $ & $T_{\rm c,0.14}$ & $R_{0.14}$ & $J_{0.14}$ & $\Omega_{0.14}$    &
$M_{0.18}$ & $\rho_{\rm c, 0.18} $ & $T_{\rm c,0.18}$ & $R_{0.18}$ & $J_{0.18}$ & $\Omega_{0.18}$  \\
\hline
    &       &   $10^8$      & $10^8$ & 0.01&  $10^{50}$ &       &   
    &   $10^8$      & $10^8$ & 0.01&  $10^{50}$ &              \\
    & \Msun &   ${\rm g/cm^3}$ &  K  & \Rsun  & erg s   & rad/s & 
      \Msun &   ${\rm g/cm^3}$ &  K  & \Rsun  & erg s   & rad/s   \\
\hline
 A1  &  1.31 & 0.42 & 0.94 & 1.10 & 1.54 & 0.97  & 1.45 & 0.56 & 0.96 & 1.01 & 1.97 & 1.26  \\
 A2  &1.34 & 0.47  & 1.01 & 1.11 &1.58 & 1.02   & 1.47 & 0.63 & 1.04 & 1.01 & 1.99 & 1.32  \\ 
 A3  &1.35 & 0.49  & 1.06 & 1.13 &1.59 & 1.03   & 1.48 & 0.66 & 1.09 & 1.03 & 2.00 & 1.34  \\ 
 A4  &1.37 & 0.54  & 1.12 & 1.15 &1.63 & 1.07   & 1.50 & 0.73 & 1.15 & 1.04 & 2.03 & 1.40  \\ 
 A5  & 1.42 & 0.78  & 1.03 & 0.93 &1.62 & 1.30  & 1.55 & 1.04 & 1.74 & 0.87 & 2.02 & 1.68  \\ 
 A6  &1.44 & 0.87  & 1.11 & 0.93 &1.63 & 1.36   & 1.56 & 1.15 & 1.15 & 0.87 & 2.01 & 1.74  \\ 
 A7  &1.46 & 0.97  & 1.17 & 0.93 &1.65 & 1.42   & 1.58 & 1.28 & 1.21 & 0.85 & 2.03 & 1.83  \\ 
 A8  & 1.48 & 1.07  & 1.23 & 0.93 &1.67 & 1.48  & 1.60 & 1.42 & 1.28 & 0.85 & 2.04 & 1.92  \\ 
 A9  & 1.55 & 1.78  & 1.18 & 0.76 &1.66 & 1.90  & 1.69 & 2.56 & 1.28 & 0.69 & 2.03 & 2.54  \\ 
 A10 & 1.55 & 1.88  & 1.26 & 0.77 &1.65 & 1.94  & 1.67 & 2.55 & 1.33 & 0.71 & 2.01 & 2.52  \\ 
 A11 & 1.57 & 2.08  & 1.33 & 0.76 &1.66 & 2.03  & 1.69 & 2.82 & 1.40 & 0.70 & 2.00 & 2.63  \\ 
 A12 &  1.58 & 2.31  & 1.40 & 0.75 &1.66 & 2.12 & 1.70 & 3.13 & 1.49 & 0.70 & 2.00 & 2.75  \\ 
\hline
 B1  & 1.36 & 0.51  & 0.96 & 1.06 &1.61 & 1.07  & 1.50 & 0.71 & 1.00 & 0.97 & 2.03 & 1.41  \\
 B2  &1.40 & 0.62  & 1.05 & 1.05 &1.66 & 1.16   & 1.54 & 0.85 & 1.09 & 0.96 & 2.07 & 1.53  \\ 
 B3  & 1.41 & 0.65  & 1.10 & 1.05 &1.67 & 1.19  & 1.55 & 0.89 & 1.14 & 0.96 & 2.07 & 1.55  \\ 
 B4  & 1.45 & 0.78  & 1.17 & 1.06 &1.71 & 1.28  & 1.58 & 1.07 & 1.22 & 0.95 & 2.09 & 1.69  \\ 
 B5  &1.46 & 0.95  & 1.06 & 0.90 &1.66 & 1.43   & 1.60 & 1.32 & 1.12 & 0.82 & 2.06 & 1.88  \\ 
 B6  &1.49 & 1.12  & 1.15 & 0.89 &1.68 & 1.54   & 1.63 & 1.56 & 1.22 & 0.81 & 2.07 & 2.02  \\ 
 B7  & 1.52 & 1.28  & 1.22 & 0.86 &1.70 & 1.63  & 1.65 & 1.79 & 1.29 & 0.81 & 2.07 & 2.15  \\ 
 B8  & 1.55 & 1.52  & 1.31 & 0.87 &1.72 & 1.76  & 1.67 & 2.12 & 1.39 & 0.77 & 2.07 & 2.32  \\ 
 B9  & 1.60 & 2.38  & 1.36 & 0.71 &1.70 & 2.19  & 1.75 & 3.94 & 1.40 & 0.62 & 2.05 & 3.13  \\ 
 B10 & 1.60 & 2.52  & 1.34 & 0.72 &1.68 & 2.23  & 1.74 & 3.81 & 1.46 & 0.64 & 2.02 & 3.05  \\ 
 B11 & 1.62 & 2.90  & 1.42 & 0.70 &1.68 & 2.38  & 1.75 & 4.32 & 1.54 & 0.64 & 2.01 & 3.22  \\ 
 B12 & 1.64 & 3.43  & 1.53 & 0.68 &1.68 & 2.56  & 1.75 & 4.91 & 1.65 & 0.61 & 1.98 & 3.42  \\ 
\hline
 C1  & 1.74 & 3.36 & 1.36 & 0.64 & 1.87 & 2.62 & 1.87 & 6.30 & 1.57 & 0.54 & 2.13 & 3.99 \\
 C2  & 1.71 & 2.66 & 1.38 & 0.69 & 1.89 & 2.35 & 1.86 & 5.02 & 1.58 & 0.59 & 2.17 & 3.58 \\
 C3  & 1.68 & 2.18 & 1.37 & 0.74 & 1.89 & 2.13 & 1.84 & 4.13 & 1.57 & 0.63 & 2.20 & 3.27 \\
 C4  & 1.64 & 1.69 & 1.35 & 0.81 & 1.88 & 1.88 & 1.81 & 3.20 & 1.54 & 0.69 & 2.23 & 2.89 \\
\hline
 D2  & 1.34 & 0.47  & 1.01 & 1.11 &1.58 & 1.02  & 1.47 & 0.63 & 1.04 & 1.01 & 1.99 & 1.32  \\
 D6  & 1.44 & 0.87  & 1.11 & 0.94 &1.63 & 1.36  & 1.56 & 1.15 & 1.15 & 0.87 & 2.01 & 1.74  \\
\hline
 T2  & 1.26 &  0.37 & 0.95 & 1.12 &1.45 & 0.90  & 1.38 & 0.47 & 0.95 & 1.05 & 1.85 & 1.14  \\
 T6  &1.39 &  0.72 & 1.02 & 0.94 &1.57 & 1.25   & 1.51 & 0.72 & 1.02 & 0.94 & 1.57 & 1.25  \\
 T10 & 1.52 &  1.57 & 1.15 & 0.76 &1.63 & 1.79  & 1.64 & 2.06 & 1.21 & 0.71 & 2.00 & 2.29  \\
\hline
\hline
\end{tabular}
\end{center}
\end{table*}

% -------------------------------------------------------------------------------------------------------------
% -------------------------------------------------------------------------------------------------------------
% -------------------------------------------------------------------------------------------------------------
% -------------------------------------------------------------------------------------------------------------
% -------------------------------------------------------------------------------------------------------------

\section{Results}\label{sect:results}

\begin{figure*}[!]
\resizebox{0.5\hsize}{!}{\includegraphics{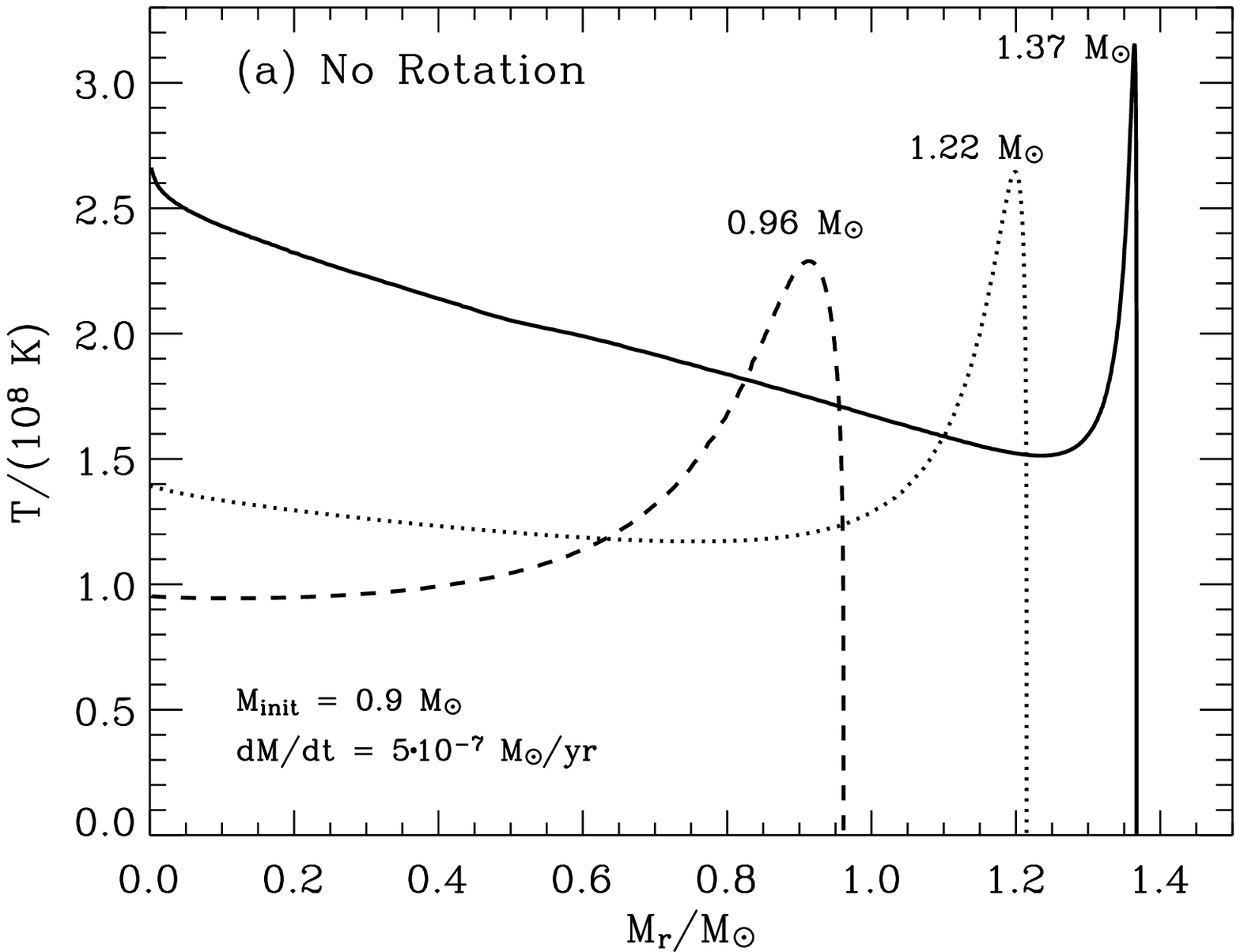}}
\resizebox{0.5\hsize}{!}{\includegraphics{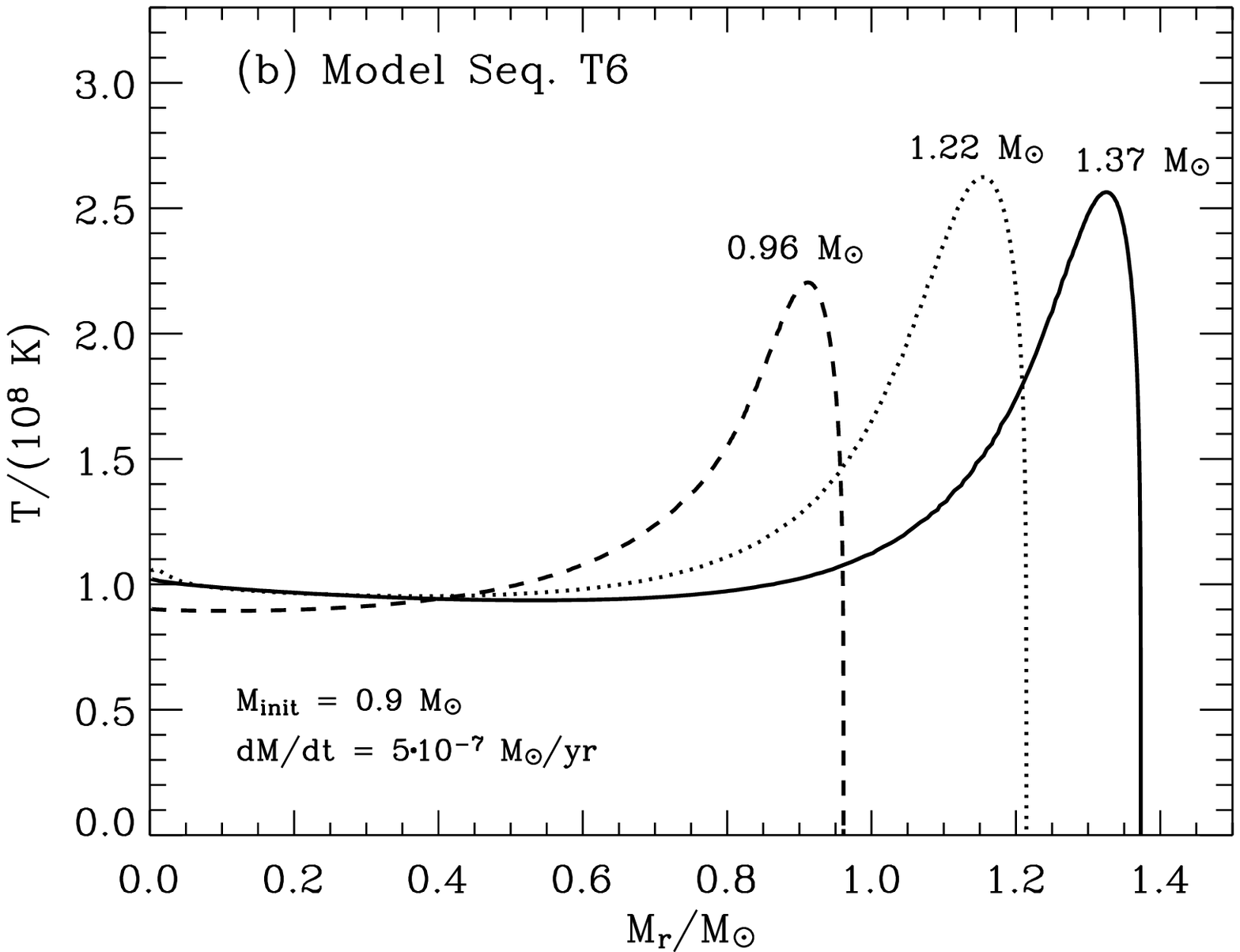}}
\resizebox{0.5\hsize}{!}{\includegraphics{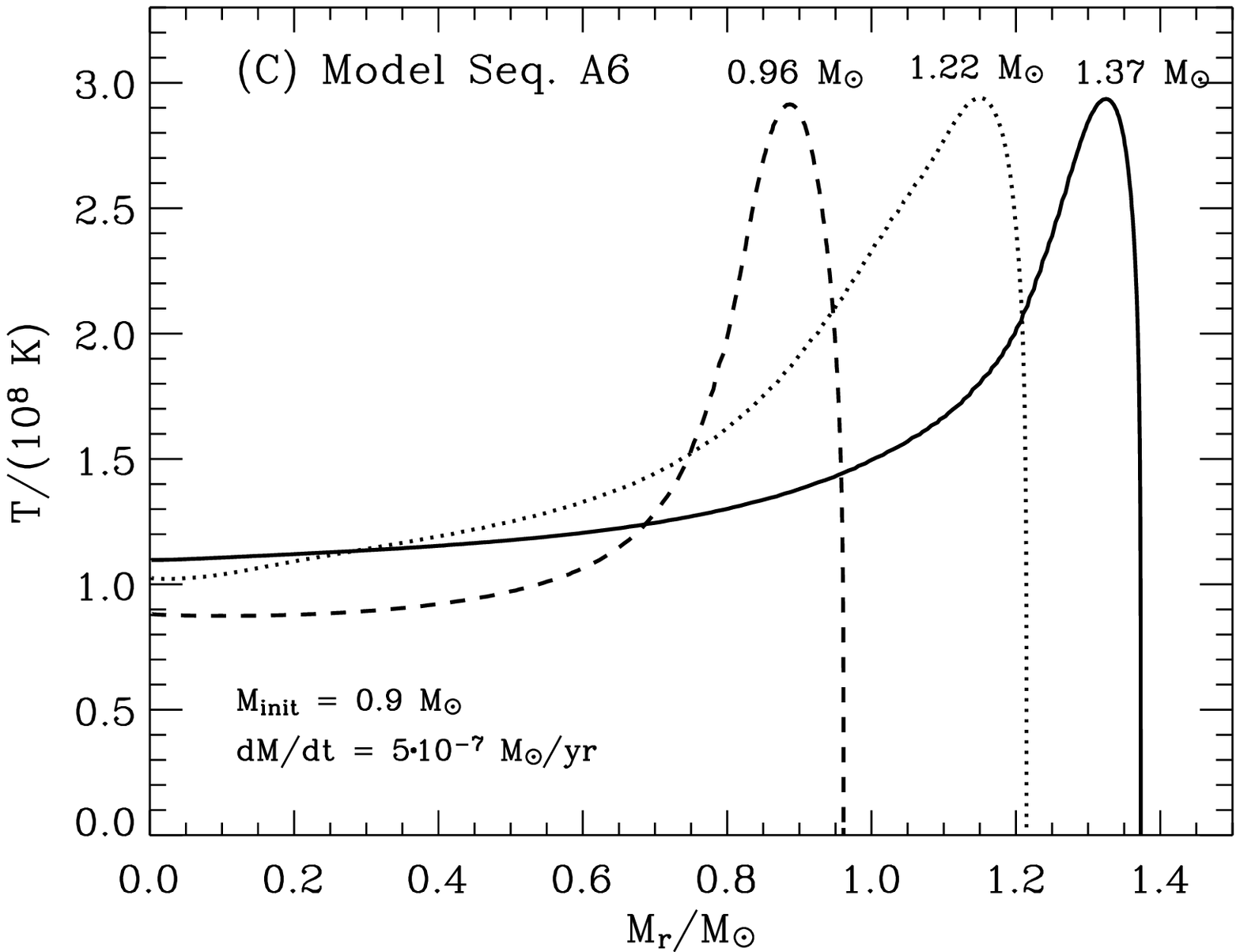}}
\resizebox{0.5\hsize}{!}{\includegraphics{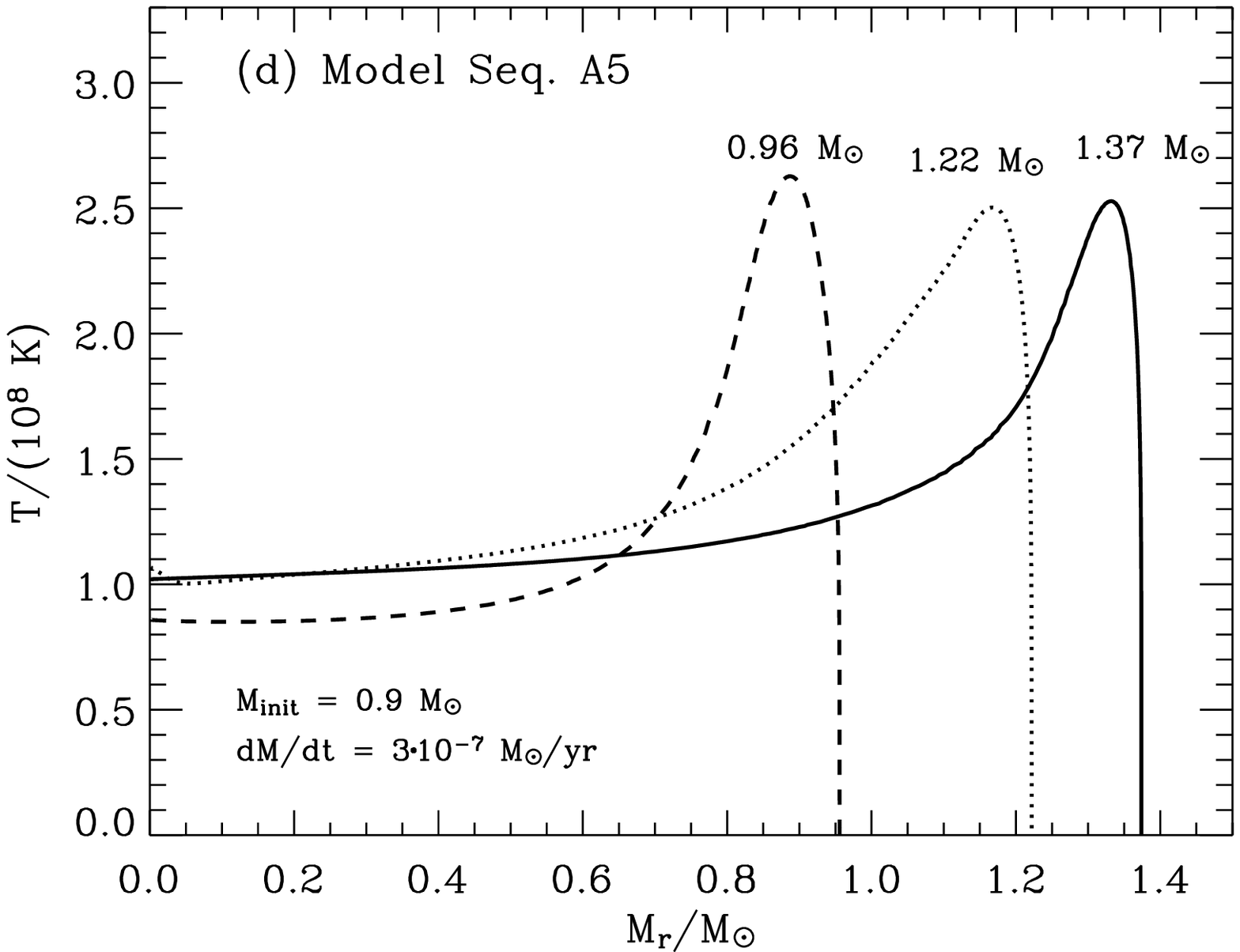}}
\caption{ 
Temperature as function of the mass coordinate 
in white dwarf models when \Mwd{} = 0.96, 1.22 and 1.37 \Msun. 
(a) non-rotating case with \Minit{} = 0.9 \Msun{} and \Mdot{} = \mdotb{},
(b) model sequence T6, 
(c) model sequence A6,
(d) model sequence A6.
}\label{fig:thermal}
\end{figure*}

In this section, we present the results of our simulations, which are
summarized in Tables~\ref{tab:result1} and~\ref{tab:result2}.
In these tables, key properties of the white dwarf models 
are given for three different epochs, i.e., when the 
ratio of the rotational energy to the gravitational potential energy
\TW{} reaches 0.10, 0.14 and 0.18. The implications of these numbers
are discussed in Sect.~\ref{sect:final}.
In the following subsections, 
after discussing the thermal evolution of our models,
we focus on the processes of the angular momentum transport
in the interior of accreting white dwarfs and its consequences
for the white dwarf structure.

\subsection{Thermal evolution}\label{sect:thermal}

The thermal evolution of non-rotating 
accreting white dwarfs has been studied by many authors 
(e.g. Iben~\cite{Iben82}; Nomoto~\cite{Nomoto82}; Sion~\cite{Sion95}; 
Townsley \& Bildsten~\cite{Townsley02}).
Piersanti et al. (~\cite{Piersanti03}) compared rigidly rotating 
white dwarf models with non-rotating ones for  
CO-accretion at various accretion rates. 
They found that a rotating white dwarf is generally cooler than its 
non-rotating counterpart, which is a 
natural consequence of the lifting effect of the centrifugal force.  
This conclusion holds also
in our differentially rotating white dwarf models (Fig.~\ref{fig:thermal}). 
However, as we shall see in Sect.~\ref{sect:spin}, 
the thermal structure in an accreting white dwarf is found to have
an interesting consequence for the redistribution of angular momentum, 
as it affects the stability criterion for the dynamical shear instability.
Therefore, before discussing the angular momentum transport, 
we examine the evolution of the thermal structure in our white dwarf 
models.

Fig.~\ref{fig:thermal} displays the temperature profiles of selected
white dwarf models of different sequences. 
The non-rotating models (Fig.~\ref{fig:thermal}a) show a rapid 
increase of the central temperature,
from $10^8$~K when \Mwd{}~=~0.96~\Msun{} to $2.7\cdot10^8$~K when \Mwd{}~=~1.37~\Msun.  
In the corresponding rotating models (Fig.~\ref{fig:thermal}b~\&~c), the
change in the central temperature 
is not significant in the considered period. 

\begin{figure}[!]
\resizebox{\hsize}{!}{\includegraphics{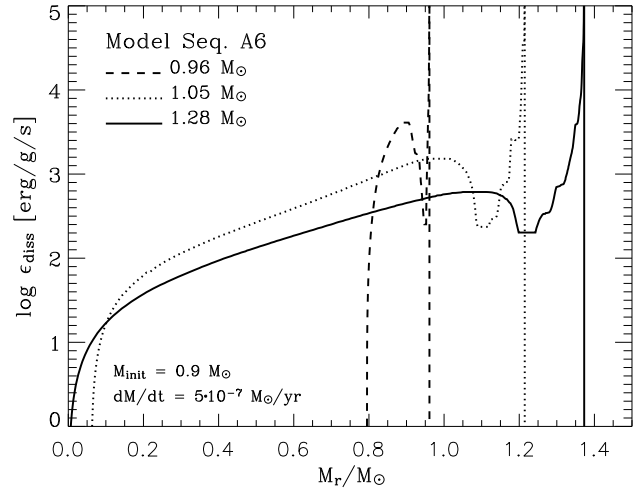}}
\caption{ 
Rotational energy dissipation rate (see Eq.~\ref{eq16}) as
function of the mass coordinate, when \Mwd{} = 0.98, 1.22 and 1.37 \Msun{}
in model sequence A6.
}\label{fig:ediss}
\end{figure}

A comparison of Fig.~\ref{fig:thermal}b and Fig.~\ref{fig:thermal}c
shows the effects of rotational energy dissipation (\Erot) described by Eq.~\ref{eq16}. 
In model sequence T6 (Fig.~\ref{fig:thermal}b), 
where  \Erot{} is neglected, 
the white dwarf models are cooler than in the corresponding models with \Erot{} considered 
(sequence A6, Fig.~\ref{fig:thermal}c). As shown in Fig.~\ref{fig:ediss}, 
the rotational energy dissipation rate can be as high as $10^3\dots10^4~{\rm erg/g/s}$
in the degenerate core, and even higher close to the surface.
The integrated energy dissipation rate 
$\dot{E}_{\rm rot} = \int \epsilon_{\rm diss} dM_{\rm r}$ reaches several times $10^2$ \Lsun{}
in these models. 
This results in a heating of the region with a strong shear motion (cf. Fig.~\ref{fig:omega} below), 
with a consequence that the temperature maxima when \Mwd{} = 0.96, 1.22 and 1.37 \Msun{} in models of sequence A6 
are higher than in the models of sequence T6.
The accretion induced heating is a sensitive function of the accretion rate: 
comparison of Fig.~\ref{fig:thermal}c with Fig.~\ref{fig:thermal}d indicates
that the white dwarf becomes significantly hotter with a higher accretion rate. 
The effects of the accretion rate and of \Erot{} for the angular momentum transport
are discussed in Sect.~\ref{sect:effmdot}.

Note that in all white dwarf models, there exists an absolute 
temperature maximum in the outer layers of the degenerate core.
Below this temperature peak, the temperature gradient $\nabla$ becomes negative
and large. This produces a strong buoyancy force in this region, 
since the Brunt-\Vaisala{} frequency is proportional to
$\nabla_{\rm ad} - \nabla$.
The temperature peak moves outward as the white dwarf accretes more mass. 
This leads to changes in the local thermodynamic condition which
plays a key role for the angular momentum
transport from the outer envelope into the inner core, as shown in the next section.

\subsection{White dwarf spin and angular momentum transport}\label{sect:spin}

\begin{figure*}[!]
\begin{center}
\resizebox{0.4\hsize}{!}{\includegraphics{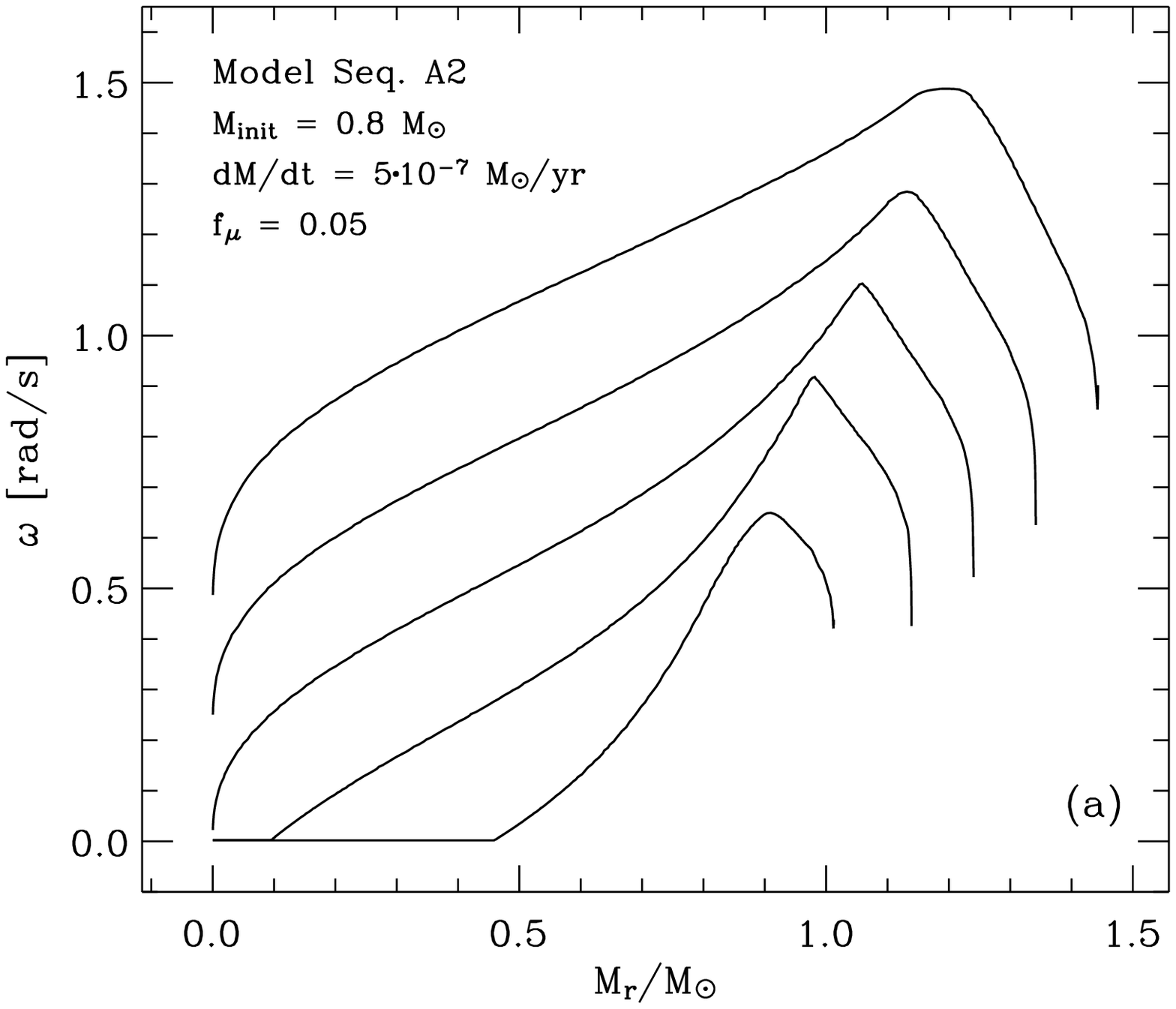}}
\resizebox{0.4\hsize}{!}{\includegraphics{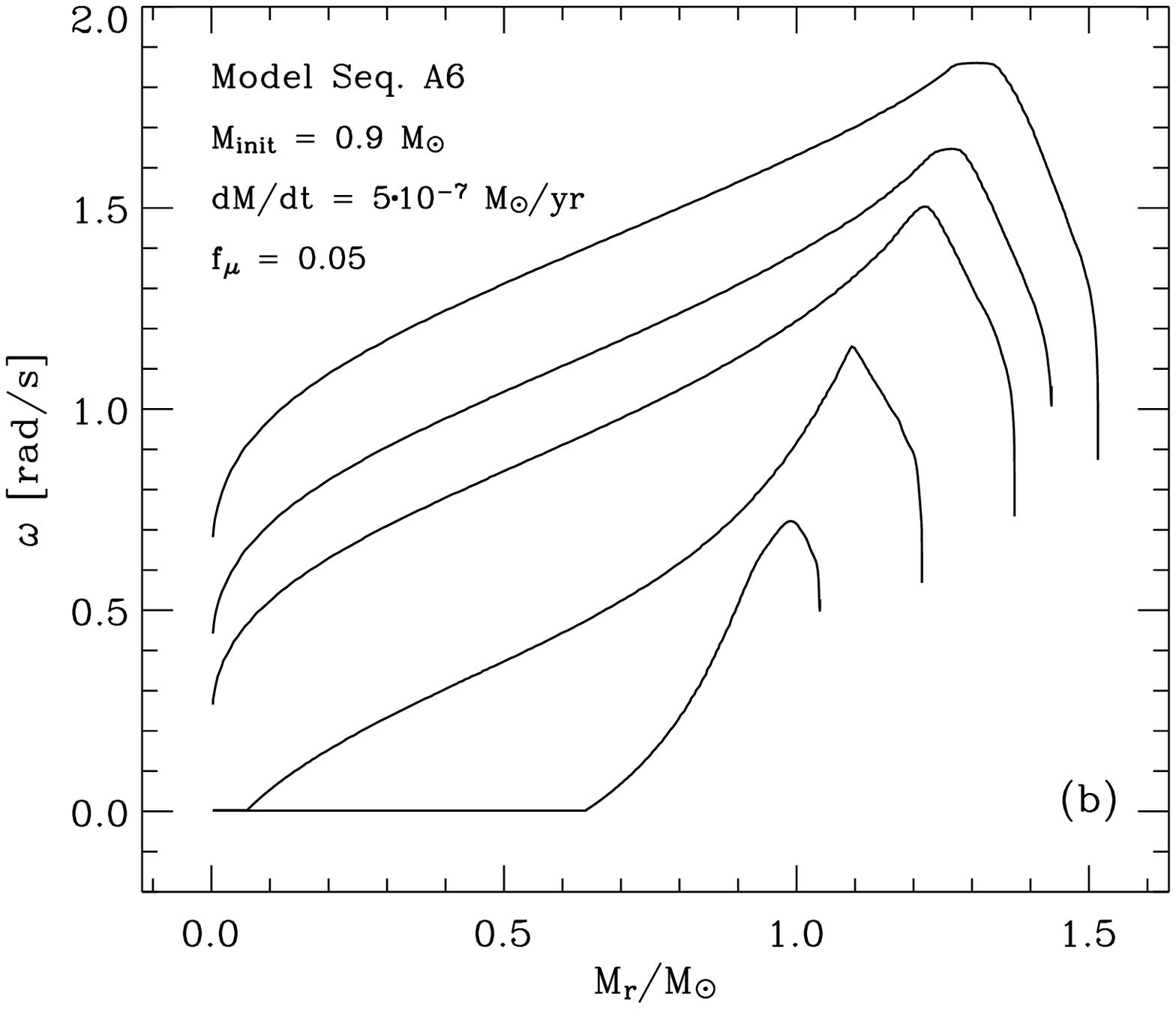}}
\resizebox{0.4\hsize}{!}{\includegraphics{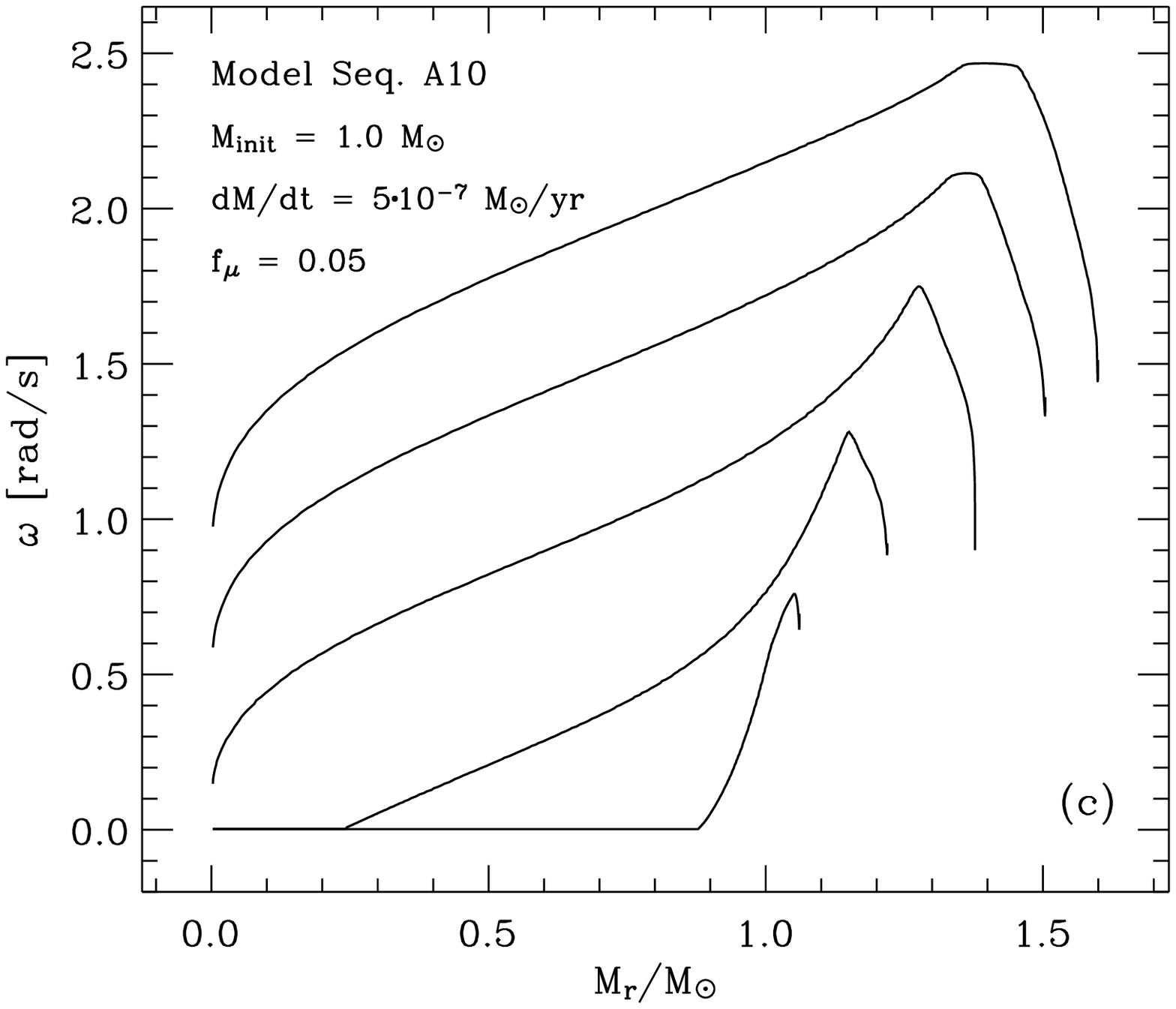}}
\resizebox{0.4\hsize}{!}{\includegraphics{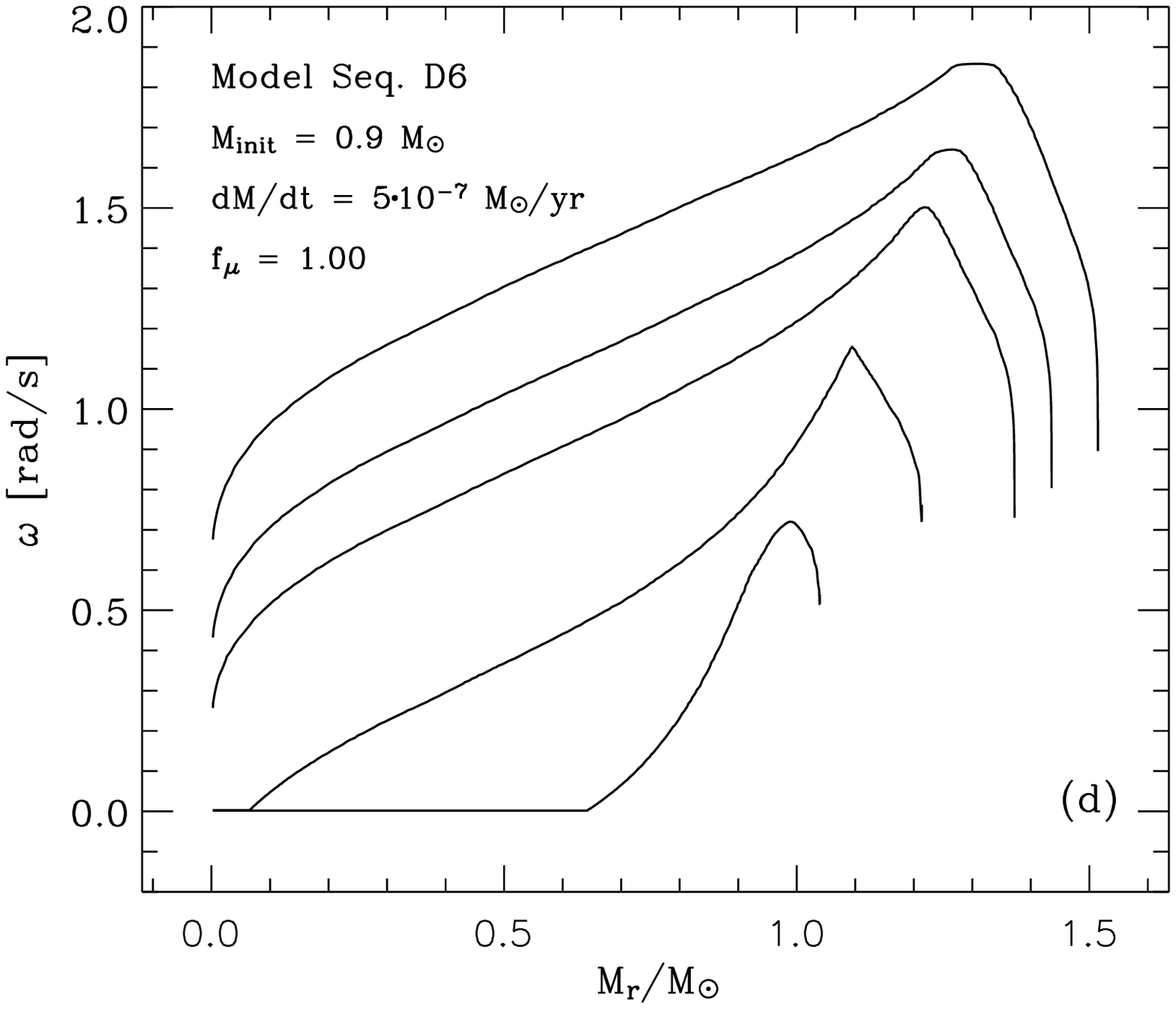}}
\end{center}
\caption{Angular velocity as function of the mass coordinate at
different white dwarf masses.
Panels (a), (b), (c) \& (d) give results for
model sequences A2, A6, A10 and D6, respectively (see Table~\ref{tab:result1}).
}\label{fig:omega}
\end{figure*}

\begin{figure}[!]
\center
\resizebox{0.9\hsize}{!}{\includegraphics{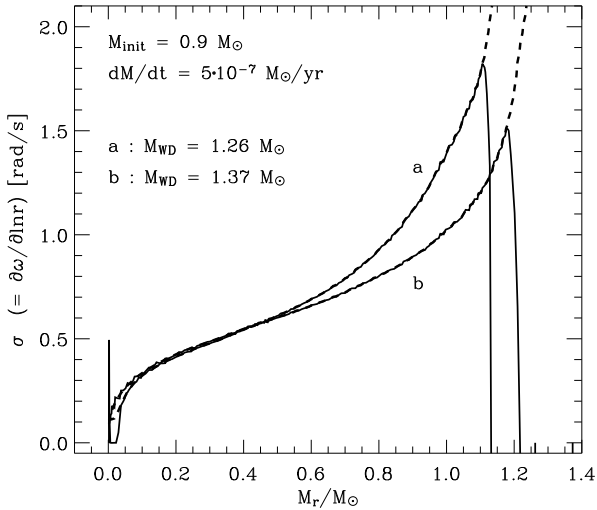}}
\caption{
Shear factor $\sigma~(= \partial \omega/\partial \ln r)$
as function of the mass coordinate
in two different white dwarf models at \Mwd{}~=~1.26~\Msun{} and 1.37~\Msun{}
from model sequence A6. 
The dashed lines denote the threshold value of $\sigma$  for 
the dynamical shear instability 
(i.e., $\sigma_{\rm DSI, crit} = (N^2/R_{\rm i, c})^{1/2}$, see Eq.~\ref{eq7}).
}\label{fig:sigma_6a}
\end{figure}

Our white dwarf models are spun up as they
gain angular momentum from the accreted matter, as described in Sect.
~\ref{sect:assumption}. Fig.~\ref{fig:omega} shows
the evolution of the angular velocity profiles
in model sequences A2, A6,  A10 and D6. 
Most strikingly,
all  white dwarf models rotate differentially, as predicted
from the discussions in Sect.~\ref{sect:angmom}. 
Note also that in every white dwarf model in Fig.~\ref{fig:omega}, 
a maximum angular velocity occurs, such that
$\sigma > 0$ in the inner core and $\sigma < 0$ in the outer layers.  
This maximum comes into existence soon after the onset of mass accretion
because 
the slowly rotating inner part contracts faster than
the rapidly rotating surface layers 
as the total mass increases.

Since angular momentum is transported in
the direction of decreasing angular velocity,
this peak in $\omega$ serves as a bottle neck
for the angular momentum transport from the outer envelope 
into the core. 
It prevents the outer envelope from slowing down efficiently by 
inward angular momentum transport. 
As a result, the surface remains to rotate close to
the critical value throughout the evolution,
severely limiting the angular momentum gain from the
accreted matter. I.e., about 60 \% of the angular momentum
of the accreted matter is rejected by the condition posed in Eq.~\ref{eq18}, and
only about 40 \% is actually retained by the time the white dwarf mass
reaches 1.4 \Msun{}, in all model sequences (see Fig.~\ref{fig:totj} below). 

Fig.~\ref{fig:sigma_6a}  shows the shear factor $\sigma$ as function
of the mass coordinate
at two different evolutionary epochs (\Mwd{} = 1.26 \& 1.37 \Msun)
in model sequence A6. The dashed line gives the threshold
value of the shear factor ($\sigma_{\rm DSI, crit}$). 
Note that the two, i.e., $\sigma$ and $\sigma_{\rm DSI, crit}$ 
converge remarkably well in the white dwarf interior, i.e.,
in \Mr{}  $\lsim$ 1.1 \Msun{} when \Mwd{} = 1.26 \Msun{} and
\Mr{} $\lsim$ 1.2 \Msun{} when \Mwd{} = 1.37 \Msun. 
This confirms the conclusion given in Sect.~\ref{sect:angmom} 
that the degenerate core of an accreting white dwarf
will rotate differentially with the shear strength 
near the threshold value
for the dynamical instability.
As already discussed in Sect.~\ref{sect:shear}, 
this is because any strong shear motion with $\sigma > \sigma_{\rm DSI, crit}$
can not be retained for a long time and should decay to $\sigma_{\rm DSI, crit}$ quickly
via the dynamical shear instability, and because
further angular momentum transport by other mechanisms
requires a longer time scale compared to the 
accretion time scale.

\begin{figure*}[!]
\resizebox{0.5\hsize}{!}{\includegraphics{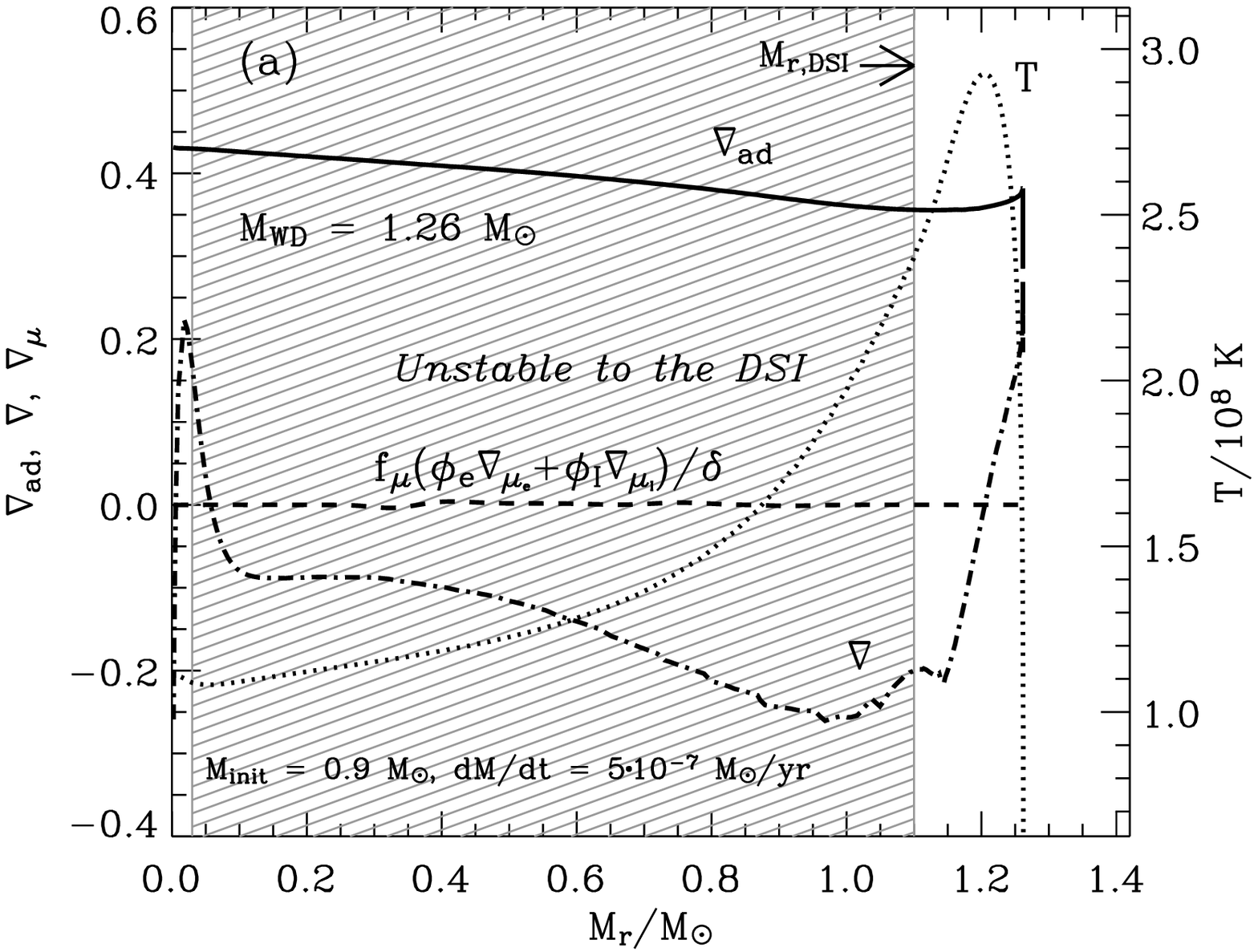}}
\resizebox{0.5\hsize}{!}{\includegraphics{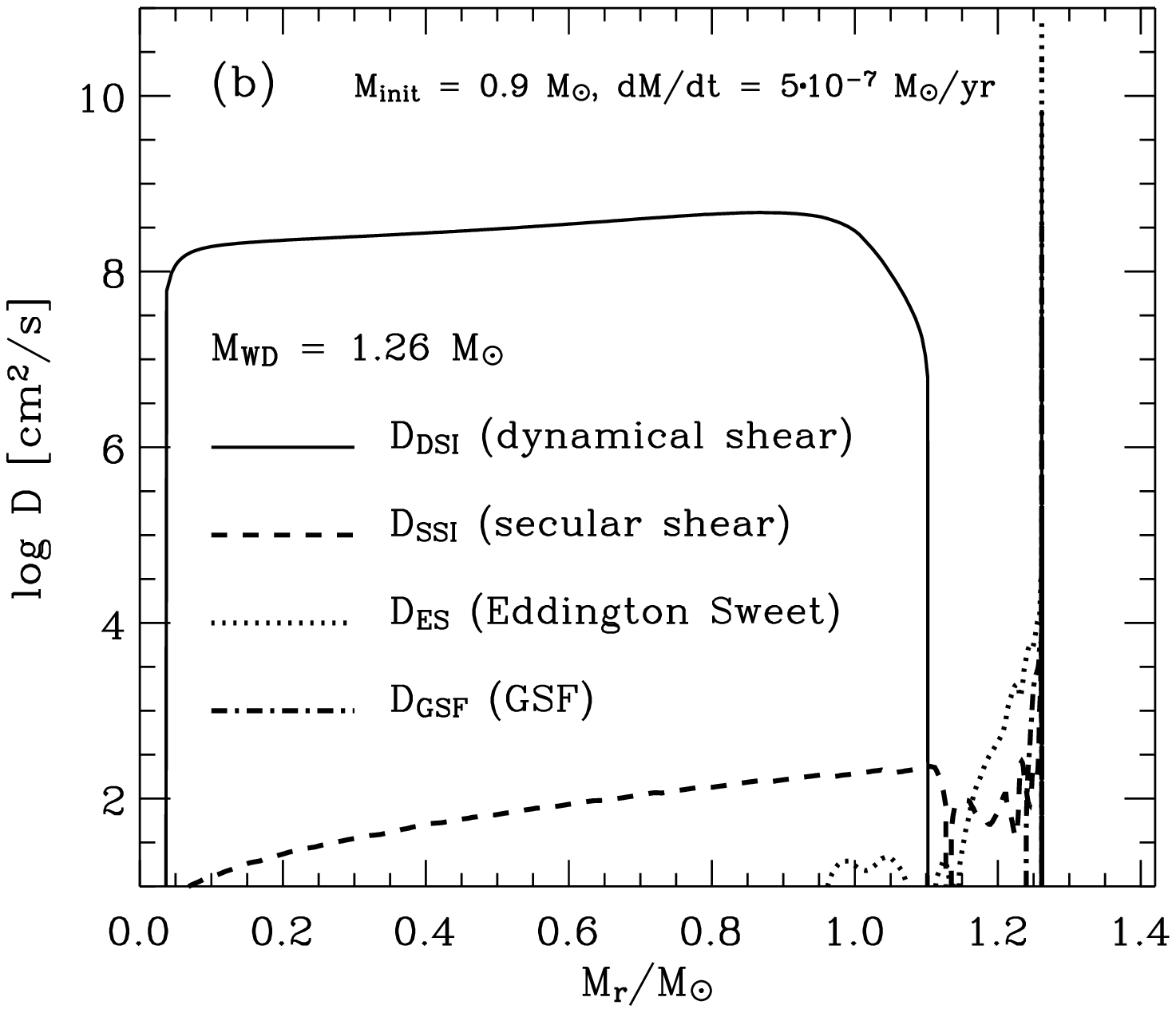}}
\resizebox{0.5\hsize}{!}{\includegraphics{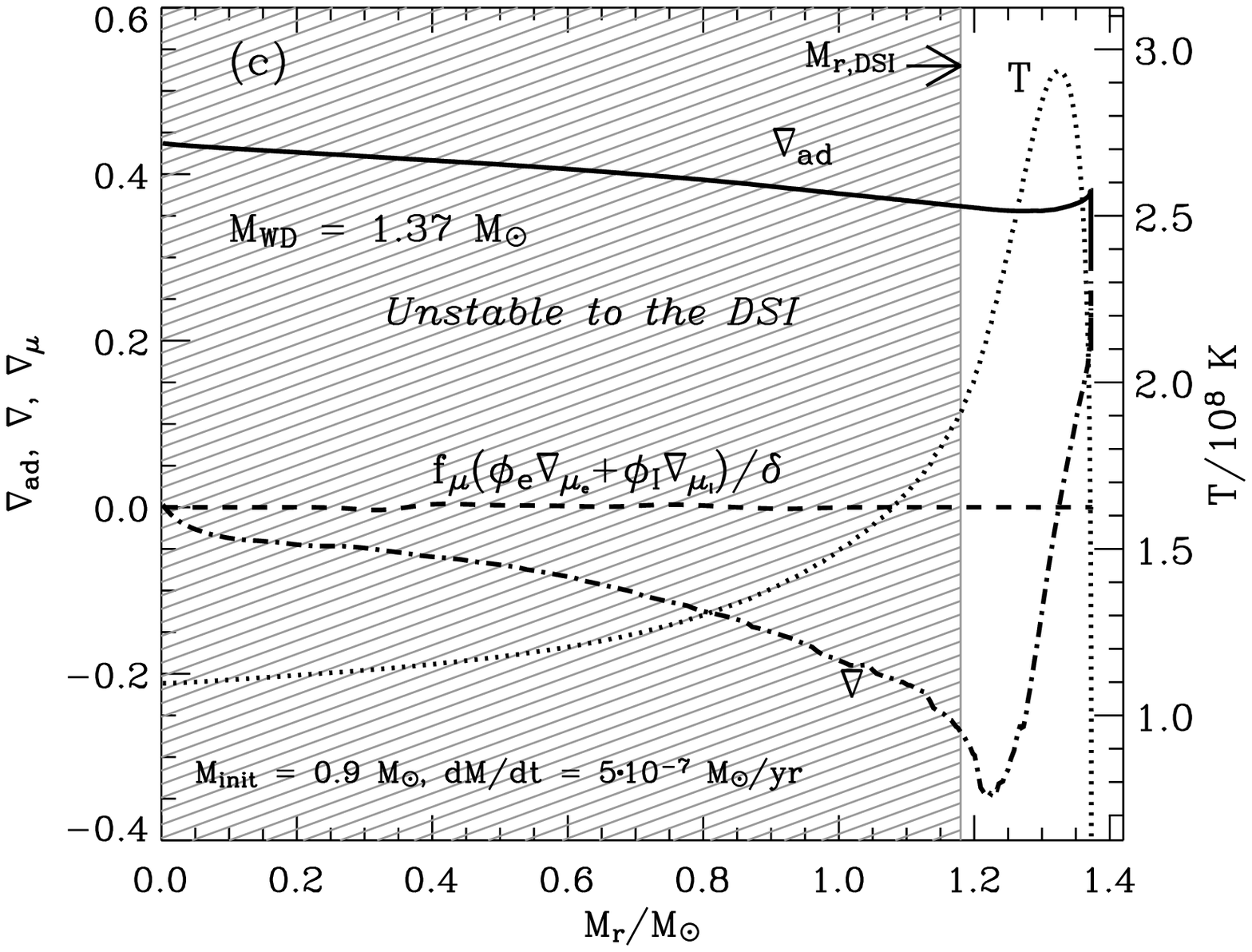}}
\resizebox{0.5\hsize}{!}{\includegraphics{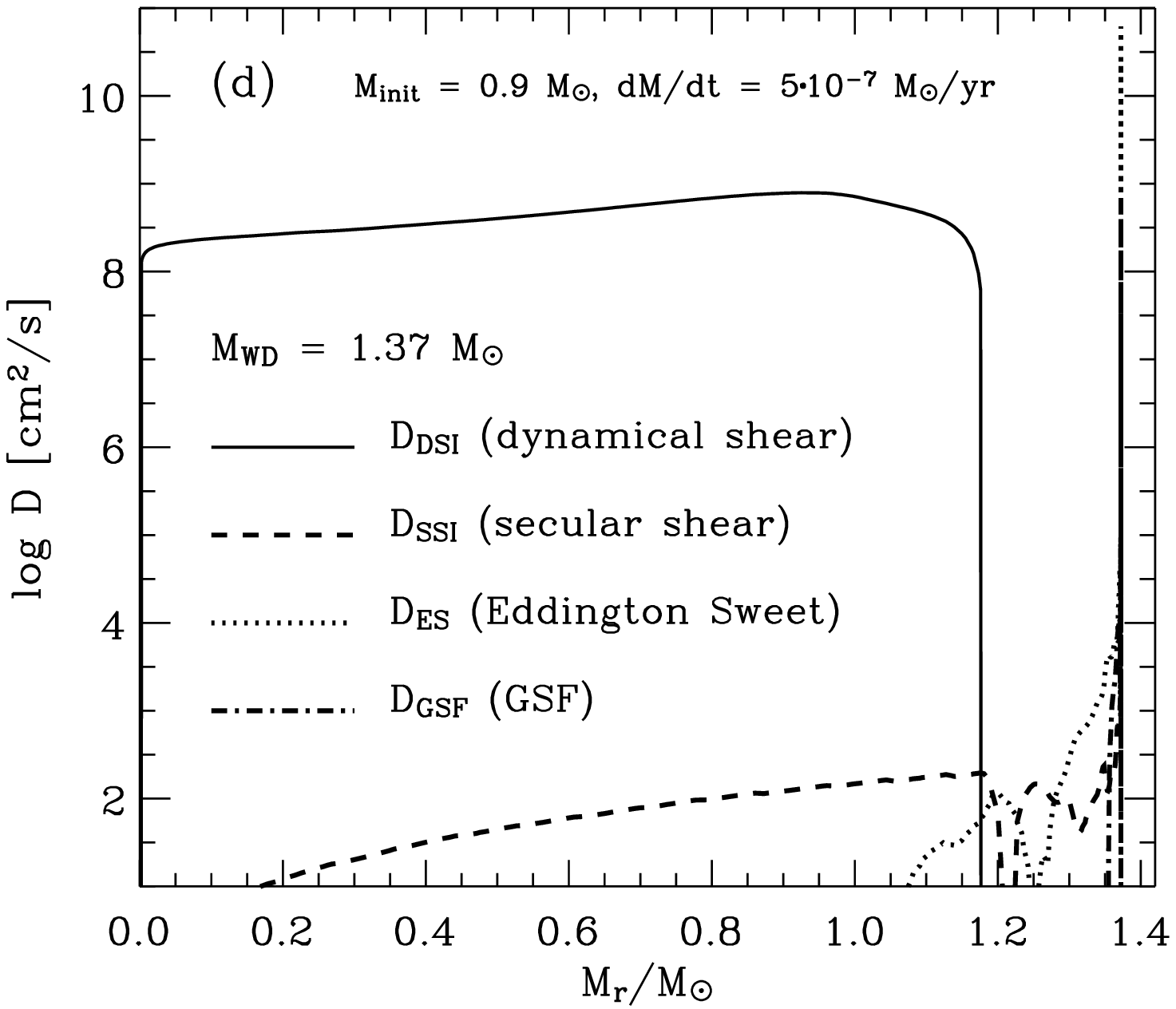}}
\caption{
(a) Thermodynamic quantities as function of the mass coordinate, 
in the white dwarf model of sequence~A6 when \Mwd{}~=~1.26~\Msun{}:
\nabad{} (solid line), $\nabla$ (dot-dashed line), $f_{\mu}$\fnabmu{} where
$f_{\mu}=0.05$ (dashed line)
and temperature (dotted line). The region where the white
dwarf is unstable to the dynamical shear instability 
is hatched.  
(b) Diffusion coefficients for rotationally induced
hydrodynamic instabilities, as function of the mass
coordinate, in the white dwarf model of sequence~A6 at \Mwd{}~= 1.26~\Msun{}.
The solid line denotes the diffusion coefficient
for the dynamical shear instability, the dashed line, the dotted line 
and the dashed-dotted lines 
are for the secular shear instability, the Eddington Sweet circulation, 
and the GSF instability respectively. 
(c) Same as in (a) but when  \Mwd{}~=~1.27~\Msun{} 
(d) Same as in (b) but when  \Mwd{}~=~1.37~\Msun{} 
}\label{fig:nabla}
\end{figure*}

Fig.~\ref{fig:nabla} depicts the situation in more detail. 
The right hand panel of this figure shows the various diffusion coefficients
as function of the mass coordinate in the white dwarf models of
sequence~A6 when \Mwd{}~=~1.26 and 1.37~\Msun{}.
This figure shows that the white dwarf consists of an 
inner dynamical shear unstable region and an outer region dominated by 
the secular shear instability and Eddington Sweet circulations.
Let us define \Mrdsi{} as the mass coordinate at the point which
divides the white dwarf into these two regions.
\Mrdsi{} changes with time not only due to the change of the shear strength, but
also due to the evolution of the thermodynamic properties as explained below. 

As shown in Sect.~\ref{sect:thermal}, the accretion induced
heating results in a temperature maximum in the outer region
of an accreting white dwarf. A steep negative temperature gradient  
(i.e., $\nabla < 0$) thus appears
just below this temperature peak. 
The buoyancy force is enhanced in this region,  
which leads to stability against the dynamical shear instability 
(See Eqs.~\ref{eq3}--\ref{eq6}). 
Therefore, the location of \Mrdsi{}
is below the region containing the strong temperature gradient,
as demonstrated in Fig.~\ref{fig:nabla}.

\begin{figure}[t]
\resizebox{\hsize}{!}{\includegraphics{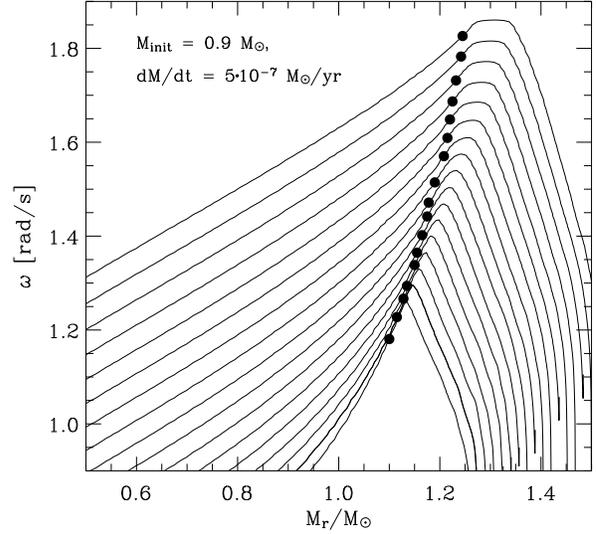}}
\caption{
Angular velocity profiles given as function of the mass coordinate
at 17 different evolutionary epochs of sequence A6:
1.26, 1.28, 1.30, 1.31, 1.33, 1.34, 1.36, 1.37, 1.38, 1.40, 1.42, 1.44, 1.45, 1.47,
1.48, 1.50, 1.52~\Msun, from the bottom to the top. 
The filled circles designate the mass coordinate $M_{\rm rm,DSI}$, 
below which the white dwarf interior is dynamical shear unstable.
}\label{fig:dynsh_crit}
\end{figure}

According to Fig.~\ref{fig:nabla}, 
\Mrdsi{} moves outward as the white dwarf mass
increases (See also Fig.~\ref{fig:sigma_6a} and Fig.~\ref{fig:dynsh_crit}). 
This outward shift of \Mrdsi{} can be understood as follows. 
As shown in Fig.~\ref{fig:sigma_6a}, the threshold value \sigdsi{}
becomes smaller near \Mrdsi{} as the white dwarf mass increases. 
Two processes contribute to this effect. 
First, the degeneracy at a given \Mr{} becomes stronger as the white 
dwarf mass increases, reducing the buoyancy force. 
Second, the region with strong temperature stratification with $\nabla < 0$ 
moves outward as indicated in Fig.~\ref{fig:nabla}a and Fig.~\ref{fig:nabla}c. 
Consequently the dynamical shear unstable region gets extended outward with 
time as shown in Fig.~\ref{fig:nabla} and 
also in Fig.~\ref{fig:dynsh_crit}.
The outward shift of \Mrdsi{} allows the angular momentum from the outer
layers to be transported inward, and thus leads to the outward shift of the
position of the maximum angular velocity as shown in Fig.~\ref{fig:dynsh_crit}.

\begin{figure}[t]
\resizebox{\hsize}{!}{\includegraphics{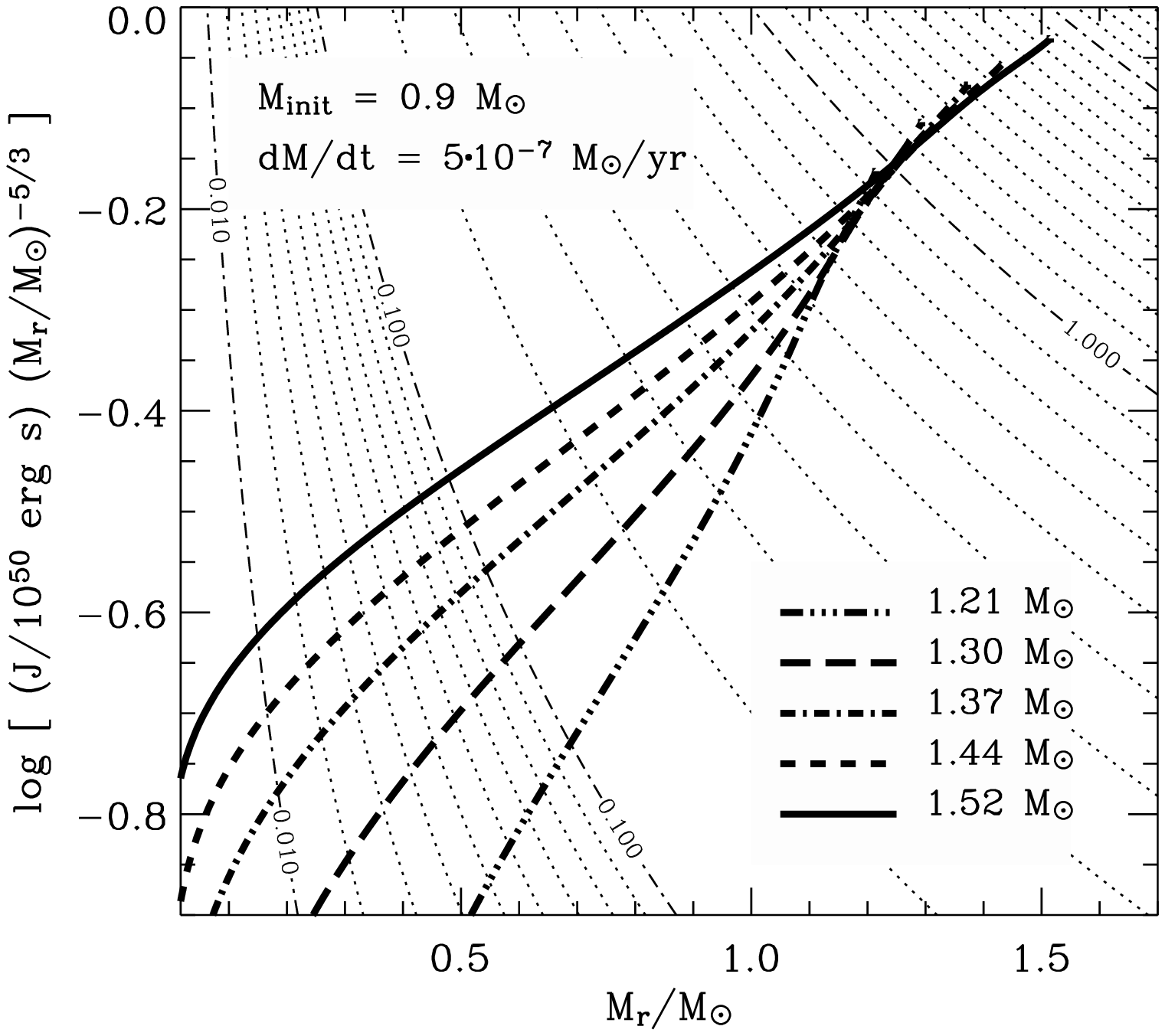}}
\caption{
Integrated angular momentum $J(M_{\rm r}) = \int^{M_{\rm r}} _0 j(m) dm$
divided by $M_{\rm r}^{5/3}$, as function of the mass coordinate at
five different evolutionary epochs of sequence A6: \Mwd{}~=~1.21, 1.30, 1.37, 1.44 and 1.52~\Msun. 
The thin contour lines denote levels of constant $J$ in logarithmic scale, 
labeled with $\log (J/10^{50} {\rm erg s})$. 
}\label{fig:jtrans}
\end{figure}

Fig.~\ref{fig:jtrans} shows how much angular momentum 
is actually transported into the white dwarf interior in model sequence A6. 
Here $J(M_{\rm r}) = \int_0^{M_{\rm r}}j(m)dm$ is the integrated angular momentum. 
In a rigidly rotating homogeneous body, we have $J(M_{\rm r})/M_r^{5/3} = const.$, 
and therefore any deviation from the zero gradient in $J(M_{\rm r})/M_r^{5/3}$
may serve as a measure of the degree of differential rotation.
A comparison of different evolutionary epochs demonstrates 
the flow of angular momentum
through the mass shells, as
$J(M_{\rm r})$ and $J(M_{\rm r})/M_{\rm r}^{5/3}$ would remain constant 
at a given mass shell
throughout the evolution if no angular momentum
were transported through this shell. Note that 
all the white dwarf models in this figure have
a similar amount of angular momentum ($J\simeq 10^{50}$~erg~s) 
at around \Mr{}~$\sim 1.2$~\Msun{},
which is close to  the angular velocity peaks shown in Fig.~\ref{fig:dynsh_crit}. 
The angular momentum transport from this point into the further interior
is shown to be efficient, due to the dynamical shear instability
as described above.

We summarize the main conclusions from this section as follows. 
a) We find accreting white dwarf models to rotate differentially 
throughout their evolution. 
b) The dynamical shear instability is the most
important mechanism for angular momentum transport in the highly
degenerate core of our white dwarf models. 
c) The angular momentum gain from the accreted matter 
and the spin-up of our accreting white dwarfs is closely related to
their restructuring and thermal evolution
as their mass increases.

\subsection{Influence of physical assumptions}\label{sect:influence}

Having understood the detailed processes of the angular momentum 
transport in an accreting white dwarf, 
let us consider the influence of the different initial 
conditions and  physical assumptions on the evolution of accreting white dwarfs.

\subsubsection{Influence of the initial white dwarf mass}\label{sect:effminit}

Tables~\ref{tab:result1} and~\ref{tab:result2} give 
the white dwarf mass when the ratio of the rotational energy to the gravitational
potential energy (\TW) in each model sequence reaches 0.1, 0.14 and 0.18 
($M_{0.1}$, $M_{0.14}$ and $M_{0.18}$). 
It is believed that a rapidly rotating white dwarf becomes secularly unstable
when \TW{}$\simeq$ 0.14
(e.g. Durisen \& Imamura~\cite{Durisen81}).
Imamura et al (\cite{Imamura95}) found that 
this critical value can be lowered to about 0.1 for strong
differential rotation (see detailed discussion in Sect.~\ref{sect:final}).

The tables  show
that the lower the initial mass,
the higher the values of $M_{0.10}$ and $M_{0.14}$ become. 
For example, $M_{0.14}$ increases from  1.34~\Msun{} to 1.55~\Msun{}
with a change of the initial mass from 0.8~\Msun{} to 1.0~\Msun{}
in the case of $f=1$ (sequences A2 and A10).
The total angular momentum
also increases at a given \TW{} with higher initial mass. 
This tendency is simply due to the fact that
more angular momentum gain is necessary
to reach a certain amount of \TW{} if a white dwarf
is more massive. 

\subsubsection{Influence of the angular momentum gain parameter $f$}

As indicated in Table.~\ref{tab:result1} and~\ref{tab:result2},  
model sequences with $f=1.0$ and those with $f=0.5$
show some differences in the results.
The adoption of $f=0.5$ yields a larger white dwarf mass at a given \TW, 
by about 0.05 -- 0.06~\Msun, 
which is a natural consequence 
of less angular momentum being accreted per unit time (Eq.~\ref{eq18}). 
This effect becomes more prominent with $f=0.3$, and $M_{0.1}$ increases
by about 24~\%  and 17~\% for \Minit{}~= 0.8~\Msun{} and \Minit{}~=~1.0~\Msun{} 
respectively, compared to the case of $f=1$.
Furthermore, the models with \Minit{}~=~1.0~\Msun{} and $f=0.3$
reach central carbon ignition when \TW{} reaches 
about 0.11. 

Interestingly, the helium accreting white dwarf models 
by Yoon et al. (\cite{Yoon04c}), where  $f=1$ is adopted, show similar \TW{} values
at a given mass as the present CO accreting white dwarf models with $f=0.3$. 
As mentioned in Sect.~\ref{sect:method}, this difference is mainly due to 
the fact that the angular momentum transport efficiency in the outermost
layers is affected by the energy generation in helium burning shell.
In particular, the Eddington Sweet circulation becomes more
efficient due to shell burning, resulting in a
more efficient outward angular momentum transport in 
the non-degenerate envelope
(cf. Fig.~\ref{fig:omega}). 
This causes a more severe restriction of the angular momentum gain from
the accreted matter, due to the condition posed by Eq.~\ref{eq18}. 
This implies that the history of the angular momentum gain 
may also be different for hydrogen accreting cases. 

However, these ambiguities concern only the actual amount of the angular
momentum gain from the accreted matter, but do not affect the history of the 
angular momentum  redistribution in the degenerate core, where
differential rotation persists during the mass accretion phase 
(Sect.~\ref{sect:shear} and~\ref{sect:spin}). In fact,
despite big differences in $f$
all model sequences show the same remarkable feature 
the carbon ignition is not reached even 
when \Mwd{}~$\gsim$~1.4~\Msun{}, due to differential rotation.

\subsubsection{Influence of accretion rate and dissipation}\label{sect:effmdot}

\begin{figure}
\resizebox{\hsize}{!}{\includegraphics{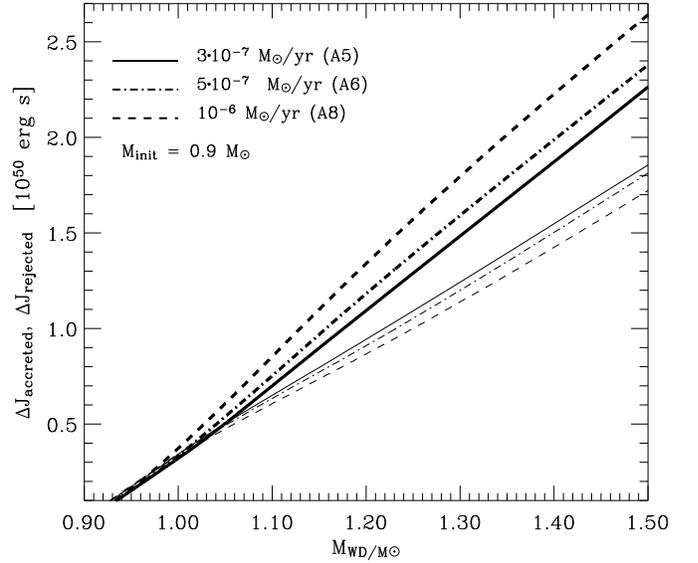}}
\caption{The evolution of the accumulated angular momentum
in model sequences A5, A6 and A10, given as function of the white dwarf mass.
The thin lines denote
the accumulated total angular momentum in the white dwarf models. 
The thick lines give the accumulated rejected angular momentum 
by Eq.~\ref{eq18}.
}\label{fig:totj}
\end{figure}

Tables~\ref{tab:result1} and~\ref{tab:result2} show that with a given initial mass, 
the white dwarf mass at at given \TW{} increases for higher accretion rate. 
Fig.~\ref{fig:totj} gives the accumulated angular momentum ($\Delta J_{\rm accreted}$) in the white dwarf, 
as well as the accumulated rejected angular momentum ($\Delta J_{\rm rejected}$) according to Eq.~\ref{eq18},
as function of the white dwarf mass, for sequences with \Minit{} = 0.9~\Msun{}, and
for three different accretion rates as indicated in the figure. 
It is shown that the higher the accretion rate, the more angular momentum is rejected
and the less angular momentum is accreted, which is the reason for the higher white
dwarf mass at a given \TW{} for a higher accretion rate. 

This accretion rate dependence can be explained by two factors. 
First, with a lower accretion rate, 
a white dwarf gains a smaller amount of angular momentum per time
and thus has more time to transport angular momentum 
into the white dwarf interior.
Second, as already pointed out in Sect.~\ref{sect:thermal}, 
a higher accretion rate results in a stronger accretion induced heating, 
which leads to higher temperatures inside the white dwarf (Fig.~\ref{fig:thermal}). 
This reduces the degeneracy in the white dwarf interior, and 
the buoyancy force  becomes accordingly stronger.
This thermal effect changes the stability condition for the dynamical shear instability
in the white dwarf interior: the higher the accretion rate, the less susceptible
to the dynamical shear instability it is, which in turn limits the angular momentum
transfer from the outer envelope into the interior more severely. 

We can also understand  why the white dwarf mass at a given \TW{}
is smaller in case \Erot{} is neglected (Table~\ref{tab:result1} and~\ref{tab:result2}) 
in terms of different thermal structures:
the white dwarf temperature becomes significantly
lower without \Erot{} as shown in Fig.~\ref{fig:thermal}, 
resulting in a lower buoyancy force in the white dwarf interior
and thus a more efficient angular momentum transport.

\subsubsection{Influence of chemical gradients}\label{sect:mugrad}	

In the model sequences discussed in the previous sections, 
the effect of $\mu$-gradients
is significantly reduced by using $f_{\mu} = 0.05$,
as described in Sect.~\ref{sect:assumption}. 
In order to understand the importance of chemical gradients for the 
angular momentum evolution, 
two model sequences (D2 and D6) are computed with $f_{\mu} = 1.0$.
The results are presented in Table~\ref{tab:result1} and~\ref{tab:result2}. 
Interestingly, it is found that the adoption of different $f_{\mu}$ 
has hardly any effect. For instance, 
we have exactly the same values for $M_{0.1}$, $M_{0.14}$ and $M_{0.18}$
in the two corresponding model sequences A2 and D2, as well as in A6 and D6.

\begin{figure}
\begin{center}
\resizebox{\hsize}{!}{\includegraphics{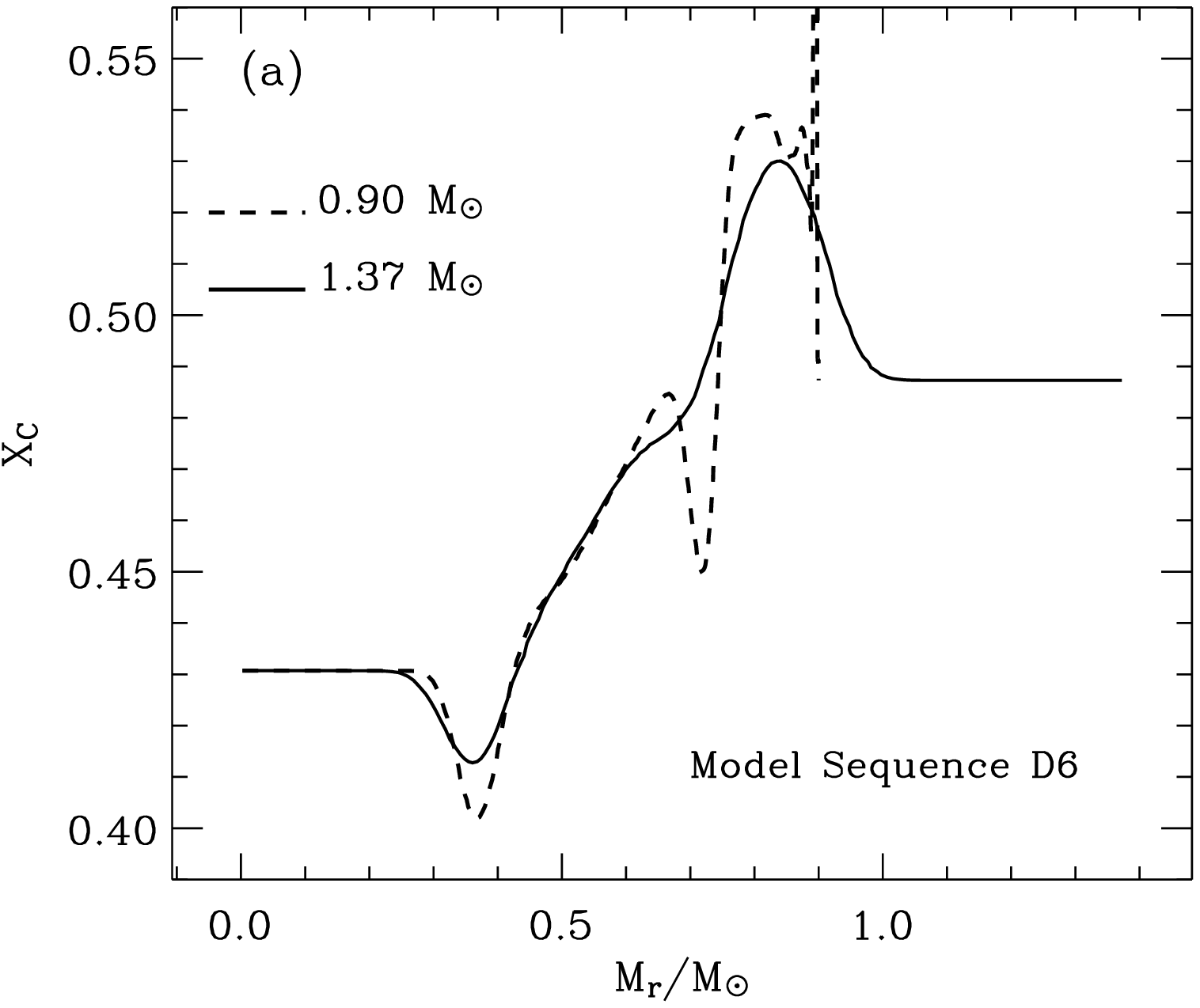}}
\resizebox{\hsize}{!}{\includegraphics{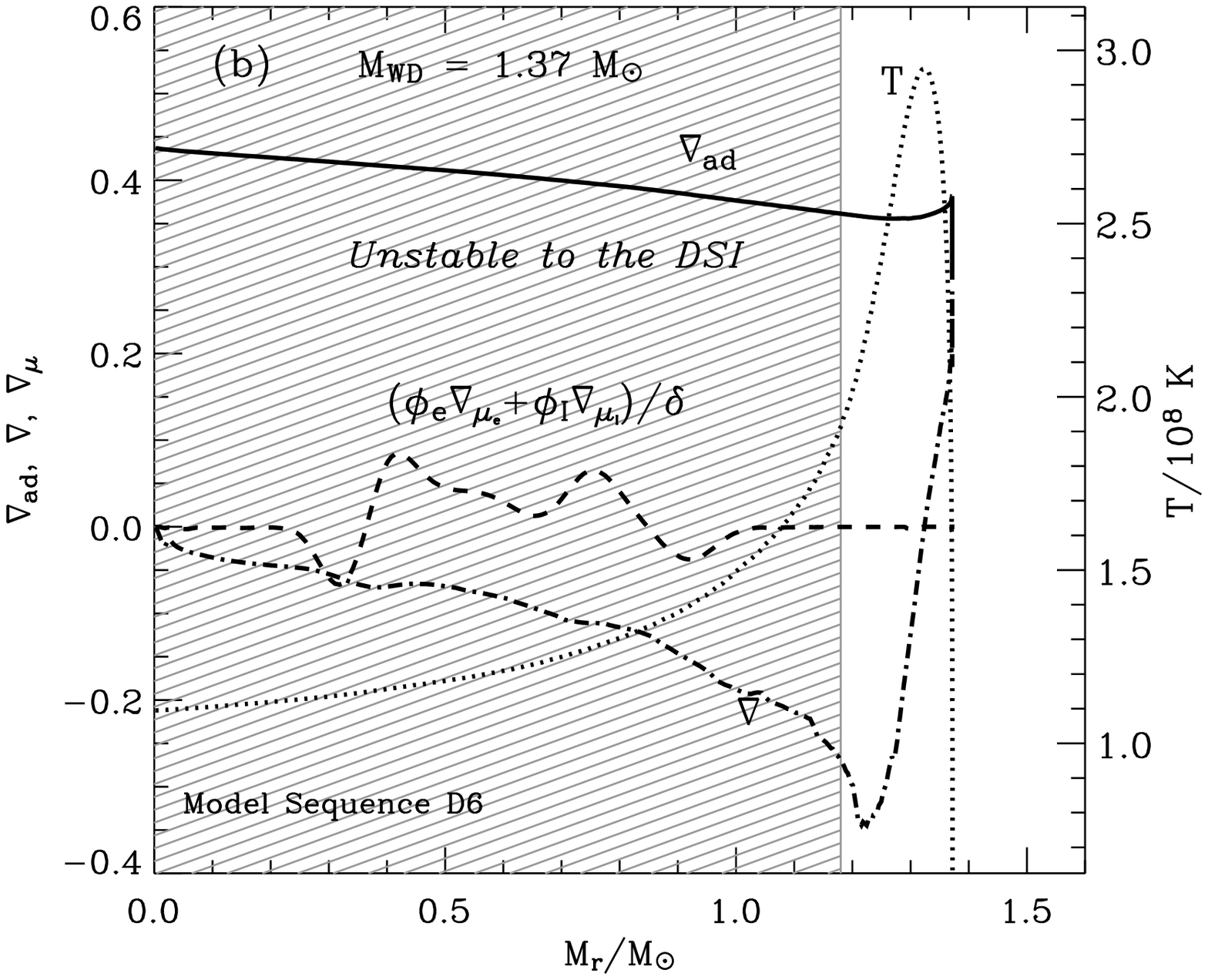}}
\end{center}
\caption{
(a) Mass fraction of carbon as function of the mass coordinate 
in two white dwarf models of sequence~D6, at \Mwd{} = 0.9 \Msun{} 
(initial model, dashed line)
and 1.37 \Msun{} (solid line).
(b) Thermodynamic quantities as function of the mass coordinate
for the same models:
\nabad{} (solid line), $\nabla$ (dot-dashed line), \fnabmu{} (dashed line)
and temperature (dotted line). The region where the white
dwarf is unstable to the dynamical shear instability 
is hatched.  
}\label{fig:mugrad}
\end{figure}

Fig.~\ref{fig:mugrad}a shows the chemical structure 
in two white dwarf models of sequence D6, in the initial model (0.9 \Msun)
and when \Mwd{} = 1.37 \Msun.
The model at 1.37 \Msun{} shows that 
although the rotationally induced 
chemical mixing smoothes out the chemical structure significantly,
a strong stratification in the chemical composition  
persists  in the range $M_{\rm r} = 0.4 - 0.8$ \Msun. 
Nevertheless, its contribution to the total buoyancy force
turns out to be too small to suppress the dynamical shear instability, 
as implied in Fig.~\ref{fig:mugrad}b. The term \fnabmu{} has a maximum value
of about 0.1, which is only 25 \% of \nabad. 

In conclusion, the effect of chemical gradients on the angular momentum transport
is negligible in the white dwarf interior, unlike in non-degenerate stars.

\subsubsection{Limitations of our 1D approximation}\label{sect:limit}

\begin{figure}
\resizebox{\hsize}{!}{\includegraphics{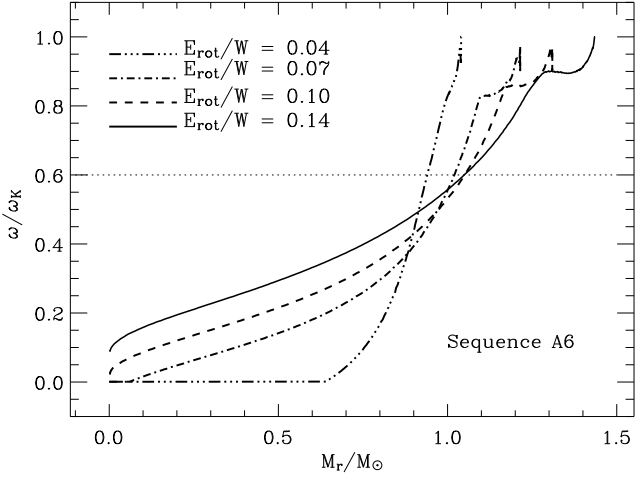}}
\caption{Angular velocity normalized to the local Keplerian value, as function 
of the mass coordinate
in white dwarf models of sequence A6, when \TW{} = 0.04, 0.07, 0.10 and 0.14
}\label{fig:wk}
\end{figure}

As pointed out in Sect.~\ref{sect:assumption}, 
our one-dimensional description of rotation 
works accurately up to about 60\% of Keplerian rotation. 
In our white dwarf models, however, the fast rotating outer layers
exceed this limit (cf. Fig.~\ref{fig:wk}). 
For instance, in model sequence A6, the outer 19 \% in mass rotate
faster than  60 \% of the Keplerian value when \TW{} = 0.1.
This fast rotating region increases to 27 \% in mass when \TW{} reaches 0.14. 
This means that our numerical models
underestimate the effect of the centrifugal force on the white dwarf structure.

The accreting white dwarf models by Durisen (\cite{Durisen77}) may be most
appropriate to evaluate 
the uncertainty due to the underestimation of the centrifugal force in the outer envelope. 
In his two dimensional models, the inner dense core with $\rho > 0.1 \rho_{\rm c}$ 
remains close to spherical, and the outer layers with $\rho < 0.1\rho_{\rm c}$
start deviating from the spherical symmetry, becoming toroidal. 
Consequently, the outer envelope is considerably extended, making the ratio
of the polar radius to the equatorial one ($R_{\rm p}/R_{\rm e}$)
as small as 0.4 when \TW{} reaches 0.1.
%This implies that the mean spin rates
%and the mean radii defined on the equipotential surfaces
%given in Table~\ref{tab:result1} and~\ref{tab:result2}
%are overestimated and underestimated, respectively. 

In our models, 
the inner slowly rotating core ($\omega < 0.6 \omega_{\rm Kepler}$)
has density $\rho \gsim 0.1 \rho_{\rm c}$ in general, and therefore
we may conclude that the inner core is accurately described in our calculations,
although it could be implicitly affected by the underestimation of the centrifugal force 
in the outer envelope. 
However, the major qualitative conclusions
of our study 
are not affected by this uncertainty: accreting white dwarfs will rotate
differentially and can grow beyond the canonical Chandrasekhar mass 
without suffering central carbon ignition, unless they lose
angular momentum through secular instabilities (Sect.~\ref{sect:final}).

\section{On the final fate of rapidly rotating massive white dwarfs}\label{sect:final}

A remarkable feature in our models is that
the white dwarfs can not reach central carbon ignition even when
their mass becomes significantly larger 
than the canonical Chandrasekhar mass of 1.4~\Msun{}
(Table~\ref{tab:result1} and~\ref{tab:result2}).
For example, in model sequence A12, the central density at \Mwd{} = 1.70 \Msun{} is 
only $3.13 \times 10^8~{\rm g/cm^3}$ (see Table~\ref{tab:result2}), 
which is still far from the carbon
ignition density (a few $10^9~{\rm g/cm^3}$ at $T \gsim 10^8$ K). 
This is in good agreement with the conclusion of previous studies of
rotating white dwarfs
(cf. Sect.~\ref{sect:previous})
that a white dwarf can be dynamically stable for masses up to $\sim 4.0$ \Msun{}
if it rotates differentially.

If a rapidly rotating massive white dwarf
would not lose angular momentum, it could not produce a Type Ia supernova.  
However, it is well known that rapidly rotating compact stars 
becomes secularly unstable to non-axisymmetric perturbations
due to the gravitational wave radiation reaction
(Chandrasekhar~\cite{Chandrasekhar70}; Friedman \& Schutz~\cite{Friedman78}), 
which is often named CFS (Chandrasekhar, Friedman and Schutz) instability.
Two representative modes are believed to be the most important ones for this 
instability. One is the bar-mode, i.e., the $f$-mode with $l=m=2$, of which the restoring
force is the buoyancy force. Here, $l$ and $m$ denote the nodal
numbers of the spherical harmonics. 
Although Friedman \& Schutz (\cite{Friedman78}) found that
$f$-modes with a higher $m$ become more susceptible to the CFS instability
compared to the case $m=2$, 
their growth time becomes usually too long to be of astrophysical interest 
(e.g. Shapiro \& Teukolsky~\cite{Shapiro83}). 
Hereafter, we will refer the CFS instability with $m=2$ as
the bar-mode instability. 
Recently it has been found that the CFS instability  can also excite
the $r$-mode, 
of which the restoring force is the Coriolis force 
(hereafter, $r$-mode instability, see Andersson \& Kokkotas~\cite{Andersson01} for a review).
Here again, the mode with $m=l=2$ is most relevant
because it gives the smallest time scale for for the growth of the instability.

In the following, 
we discuss the importance of these instabilities for our
white dwarf models, and 
derive implications 
for the final fate of rapidly rotating massive white dwarfs.

\subsection{Bar-mode instability }\label{sect:barmode}

The bar-mode instability can operate when \TW{} exceeds a certain critical value [\TWc{}]
in a rotating star, as studied by many authors 
(Ostriker \& Tassoul~\cite{Ostriker69}; Ostriker \& Bodenheimer~\cite{Ostriker73}; 
Durisen~\cite{Durisen75b},~\cite{Durisen77}; 
Bardeen et al.~\cite{Bardeen77}; Durisen \& Imamura~\cite{Durisen81}).
Although \TWc{} is found to be about 0.14 for 
a wide range of rotation laws
and equations of states (e.g. Durisen~\cite{Durisen75b}; Karino \& Eriguchi~\cite{Karino02}), 
Imamura et al. (\cite{Imamura95}) showed that
it tends to decrease for strong differential rotation.
In particular, we note that the rotation law in our models
bears a similarity with  one of their rotating polytrope models
with $n'=\infty$ (see Imamura et al. for the definition of $n'$), 
in which the spin rate shows a maximum where
the gradient in $\omega$ changes its sign, as in our models (Fig.~\ref{fig:omega}). 
In these models, \TWc{} turns out to be as small as 0.09, which 
is significantly smaller than the canonical value of 0.14.

According to Chandrasekhar (\cite{Chandrasekhar70}) and Friedman \& Schutz (\cite{Friedman75}),
the  growth time of the bar mode instability
in Maclaurin spheroids is given by
\begin{equation}\label{eq:taubar}
\tau_{\rm bar} \sim 10^{-8} \frac{R}{c}\left(\frac{R\Omega}{c}\right)^{-6}
\left[\frac{E_{\rm rot}}{|W|} - \left(\frac{E_{\rm rot}}{|W|}\right)_{\rm bar, c}\right]^{-5} ~{\rm sec,} 
\end{equation}
for $0 < E_{\rm rot}/|W| - (E_{\rm rot}/|W|)_{\rm c} \ll 1$,
where $\Omega$ is the mean angular velocity and $R$ is the radius.

Since we have $E_{\rm rot}\simeq k^2 MR^2\Omega^2$ and $|W| \simeq GM^2/R$ 
(where $k$ is the 
dimensionless radius of gyration), the growth time of the instability
can be rewritten as (cf. Friedman and Schutz~\cite{Friedman75}; Hayashi et al.~\cite{Hayashi98})
\begin{align}\label{eq:taubar2}
\tau_{\rm bar} \sim  1.6 & \left(\frac{R}{0.01 {\rm R_{\odot}}}\right)^4
\left(\frac{M}{{\rm M_{\odot}}}\right)^{-3}
\left(\frac{E_{\rm rot}}{|W|}\right)^{-3} \nonumber \\
&\times \left[\frac{E_{\rm rot}}{|W|} - \left(\frac{E_{\rm rot}}{|W|}\right)_{\rm c}\right]^{-5} ~{\rm sec,}
\end{align}
where $k^2 = 0.4$ has been used. 
Although our white dwarf models deviate significantly from the Maclaurin
spheroids, this approximation may give an order of magnitude estimate
for the growth time scale of the bar-mode instability. 

\begin{figure}
\resizebox{\hsize}{!}{\includegraphics{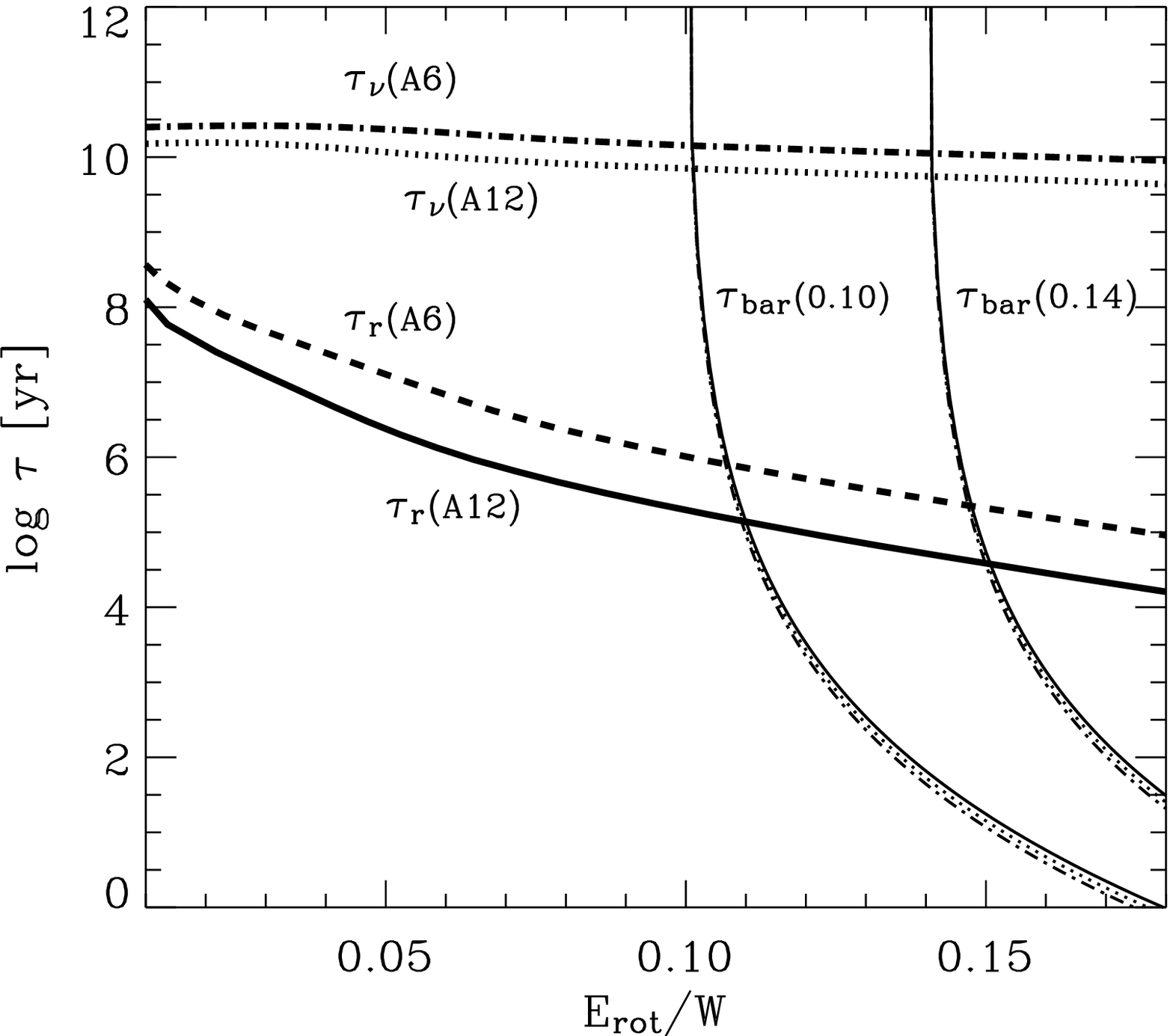}}
\caption{
Growth time scales for the bar mode instability 
($\tau_{\rm bar}$, Eq.~\ref{eq:taubar2})
as well as for the r-mode instability ($\tau_{\rm r}$, Eq.~\ref{eq:taur}). 
$\tau_{\rm bar}$ is
plotted as function of \TW, with thin lines
for 3 different masses: 1.4 \Msun{} (solid line),
1.5 \Msun{} (dotted line) and 1.6 \Msun{} (dashed dotted line).  
Two different values of \TWc{} (0.1 and 0.14) for the bar-mode instability 
are considered as indicated in the figure.  
The thick dashed line denotes $\tau_{\rm r}$
in the models of sequence A6, while the thick solid line is for sequence A12.
The thick dashed-dotted and dotted lines give  the viscous time scale
for sequences A6  and A12 respectively.
}\label{fig:tau}
\end{figure}

Fig.~\ref{fig:tau} shows $\tau_{\rm bar}$ as function of \TW{} 
for four different masses as indicated in the figure caption. 
Here, $R  = 0.01$ \Rsun{}
is assumed since the mean radii of our white dwarf models do not 
differ much from 0.01 \Rsun{} for all \TW{} values. 
Two different values for \TWc{}, i.e.,  0.1 and 0.14 are considered
in the figure. 
The growth time scale of the bar-mode
instability is a sensitive function of \TW, while
it does not change much for different white dwarf masses:
In case of \TWc{} = 0.10, $\tau_{\rm bar}$ is as large as $10^{10}$ yr at \TW $\simeq$ \TWc,
but it drops to $10^4$ yr when \TW{} $\simeq$ 0.12.
Furthermore, $\tau_{\rm bar}$ becomes only about 10 yrs when \TW{} approaches 0.15. 

Durisen (\cite{Durisen77}) found that the angular momentum loss time scale 
is comparable to the growth time scale of the instability 
($\tau_{J, {\rm bar}} \simeq \tau_{\rm bar}$), 
and therefore
we may conclude that the bar-mode instability can be 
an efficient mechanism to remove angular momentum
for white dwarfs with \TW{}$>$ \TWc, 
as long as this is not suppressed by turbulent motions 
(see Sect.~\ref{sect:further}).

\subsection{$r$-mode instability}\label{sect:rmode}

Recently, Andersson (\cite{Andersson98}) and  Friedman \& Morsink (\cite{Friedman98}) have
found that $r$-modes in all rotating inviscid stars 
are unstable to the CFS instability.  
The possibility of gravitational wave radiation from
fast rotating white dwarfs due to the $r$-mode instability 
was also discussed
by Hiscock (\cite{Hiscock98}), Andersson et al. (\cite{Andersson99}) and Lindblom (\cite{Lindblom99}). 
Lindblom (\cite{Lindblom99}) gives the growth time scale for this instability with $m=l=2$ as
\begin{equation}\label{eq:taur}
\frac{1}{\tau_{\rm r}} =\frac{2\pi}{25}\left(\frac{4}{3}\right)^8
\frac{G}{c}\int_{0}^{R} \rho \left(\frac{r\Omega}{c}\right)^6 dr.
\end{equation}
This time scale is plotted in Fig.~\ref{fig:tau} 
for two model sequences (A6 and A12). 
Interestingly, we find that the $r$-mode instability 
becomes important when \TW{}$\lsim$ \TWc, compared to the bar-mode instability.
The sensitivity of the time scale  $\tau_{\rm r}$ to \TW{}  is much weaker than
that of $\tau_{\rm bar}$, and  $\tau_{\rm r}$ remains in the range $10^5 \dots 10^8$ yr
for \TW $<$ \TWc{}. 
Note that $\tau_{\rm r}$ in our models has much smaller values 
than in Lindblom (\cite{Lindblom99}),  
who found $\tau_{\rm r}$ to be larger than $6\times10^9$ yr 
in the observed DQ Her objects.
This difference is mainly due to the fact that our white dwarf 
models are rotating differentially, with the outer layer rotating faster 
than the inner core.
More precise estimates of $\tau_{\rm r}$ require, of course, a 
multidimensional study, which is beyond 
the scope of this paper.
Nevertheless, the present estimates indicate
that the $r$-mode instability might be a promising mechanism
for the removal of angular momentum 
in a differentially rotating massive white dwarf  
within an interestingly short time scale ($\lsim 10^8$ yr), 
when  \TW{} $<$ \TWc{}

\subsection{Implications for the final fate of accreting white dwarfs}\label{sect:further}

The previous sections imply that
our differentially rotating massive white dwarfs can lose
angular momentum 
via the bar-mode instability if \TW{} $\gsim$ \TWc{},
and via the $r$-mode instability if \TW{} $\lsim$ \TWc. 
These instabilities would remove 
angular momentum from the outer layers of the white dwarfs ---
as shown in the numerical simulations of rotating neutron stars (Lindblom et al.~\cite{Lindblom02}) --- where 
most of the white dwarf angular momentum is located in our models. 

Although the CFS instability is likely suppressed 
in the presence of strong viscosity for both, the r-mode and the bar-mode 
(e.g. Andersson \& Kokkotas~\cite{Andersson01}), 
Imamura et al. (\cite{Imamura95}) suggest that secularly unstable modes may not be
necessarily damped in the presence of a large effective viscosity due to 
turbulence.
Given this uncertainty, we consider both cases, i.e., we assume a
strong damping of secular instabilities due to turbulence
induced by the shear instability as Case~I,
and a persistence of secular modes even when the shear
instability is present, where
the CFS instability is only 
affected by the microscopic viscosity (Case~II).

\subsubsection{Case I}

If the unstable CFS modes are damped by turbulence, 
an accreting white dwarf will not experience any angular momentum loss due to
the CFS instability during the mass accretion phase in which
the dynamical and secular shear instability 
persist throughout the white dwarf interior 
(see Sect.~\ref{sect:spin}), 
no matter how large \TW{} becomes.
Once the mass accretion ceases, 
the angular momentum redistribution
in the white dwarf will continue until
the degree of the differential rotation becomes 
weak enough for the shear instability to disappear. 
Once the fluid motion in the white dwarf becomes laminar, 
only the electron and ion viscosities
will contribute to the viscous friction.

The time scale of viscous dissipation ($\tau_{\nu}$) through the electron 
and ion viscosity 
is plotted in Fig.~\ref{fig:tau} for model sequences A6 and A10. 
Here $\tau_{\nu}$ is calculated following the equation (4.2) in Lindblom (\cite{Lindblom99}). 
It is found that $\tau_{\nu}$ is far larger than $\tau_{\rm r}$
and $\tau_{\rm bar}$ (except for $\tau_{\rm bar}$ at \TW{} $\simeq$ \TWc).
This implies that the unstable modes of the CFS instability 
will not be damped by the microscopic viscosity
before they grow to a dynamically meaningful level, once
the shear instability has decayed.

If the white dwarf mass has grown to \Mwd $\gsim$ 1.4 \Msun{}
by the end of the mass accretion phase,
carbon ignition at the white dwarf center
will be delayed as follows.
After the mass accretion stops, the white dwarf will evolve without losing
angular momentum until the shear instability decays.
When the white dwarf interior becomes laminar, 
the white dwarf will begin to lose the angular momentum 
either by the bar-mode instability if \TW{} $>$ \TWc{}, or 
by the $r$-mode instability if \TW{} $\lsim$ \TWc{}. 
The core density will increase as it loses angular momentum
and carbon will ignite at the center, eventually. 

Can such a white dwarf end in a Type Ia supernova? 
It depends on 
how much time is spent from the halt of mass accretion to carbon ignition, 
the time scale to which we will refer as $\tau_{\rm delay}$. 
Nomoto \& Kondo (\cite{Nomoto91}) found
that if the core of a white dwarf has solidified,
core carbon ignition is likely to end in a collapse
rather than an explosion. I.e., to obtain a SN~Ia,
carbon ignition should occur
before the white dwarf core is crystallized.

\begin{figure}
\resizebox{\hsize}{!}{\includegraphics{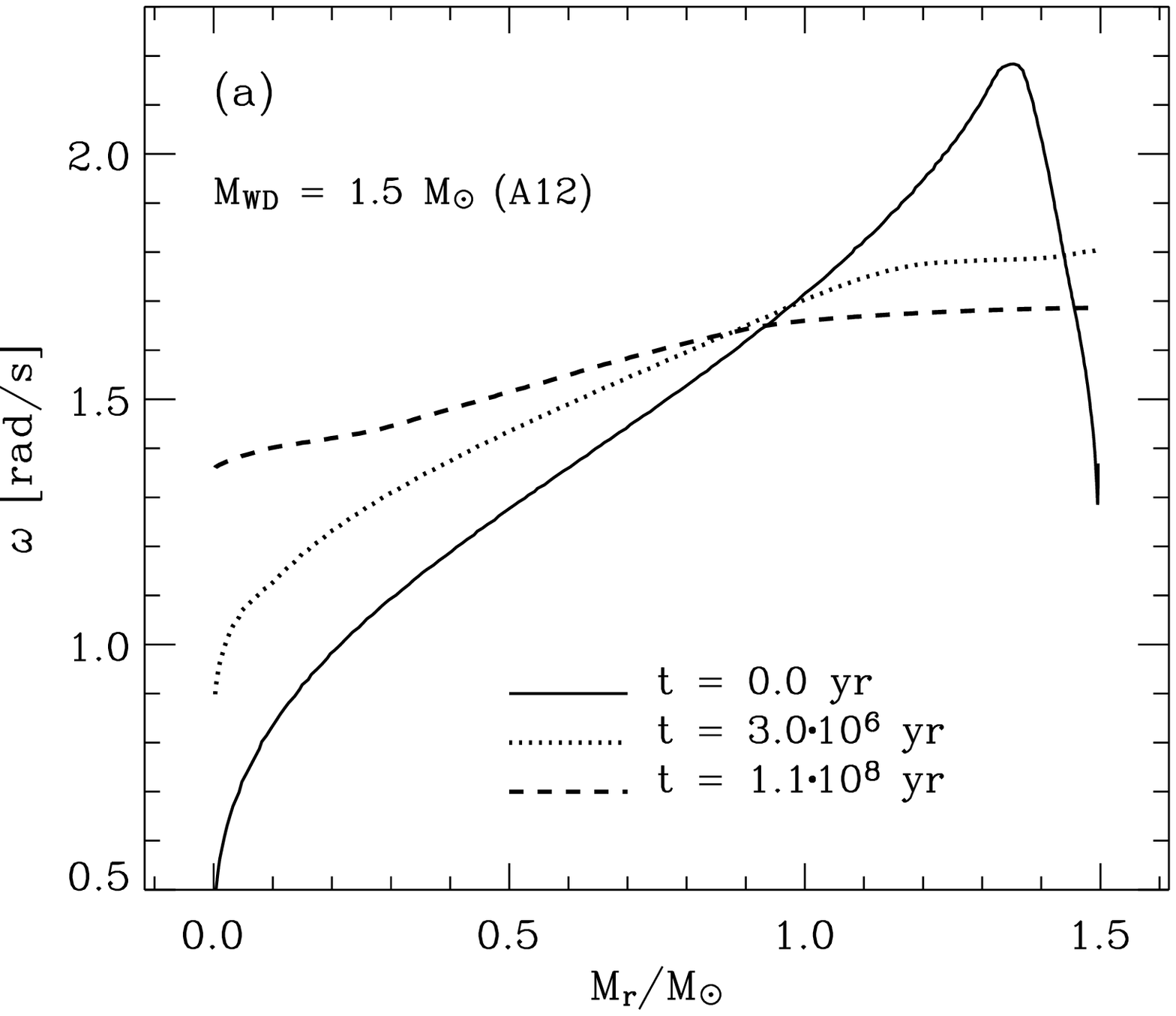}}
\resizebox{\hsize}{!}{\includegraphics{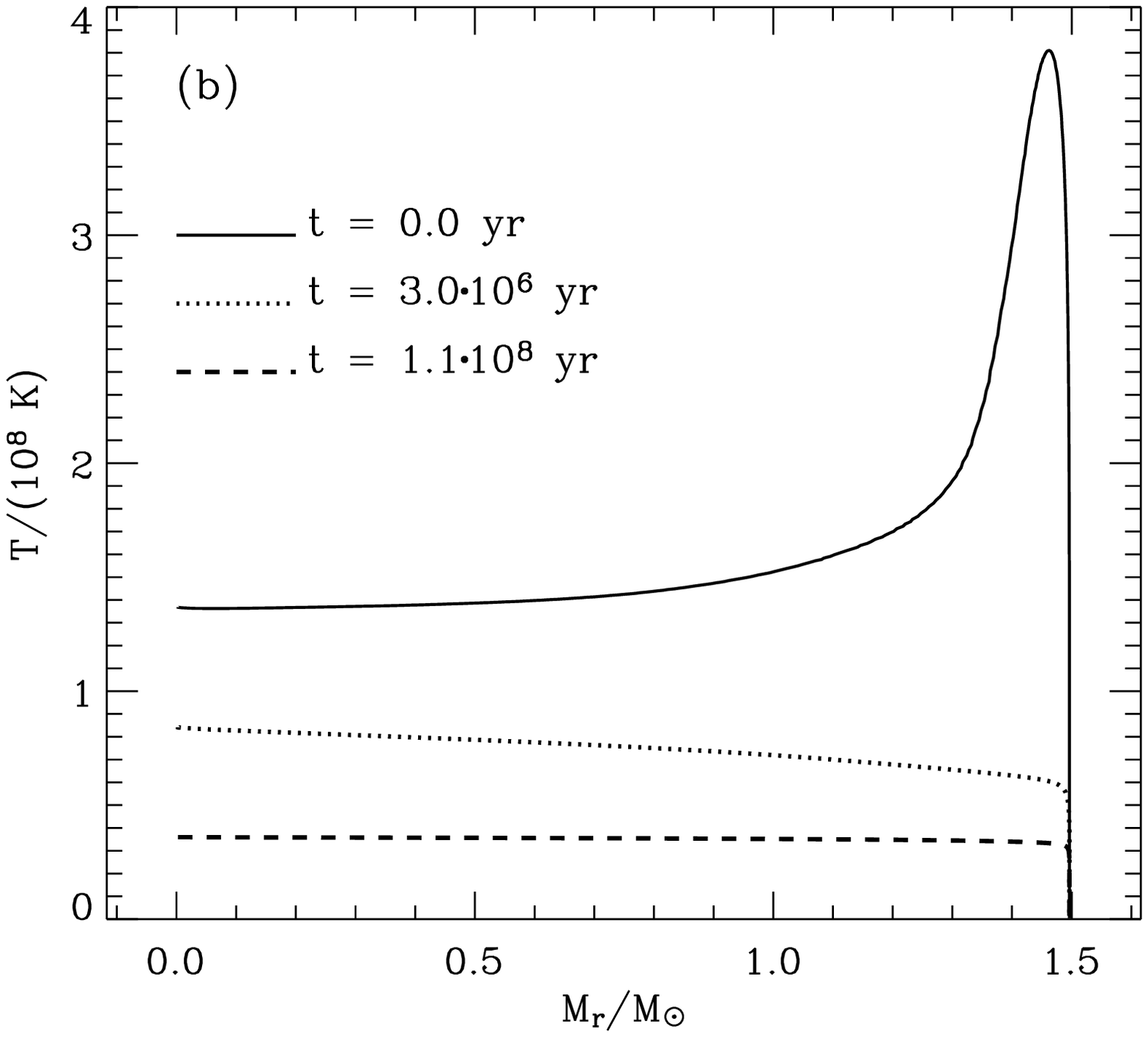}}
\caption{
Evolution of the angular velocity in (a), and 
temperature in (b), of a rapidly rotating 1.50 \Msun{} white dwarf 
after the mass accretion stops.
The initial model at $t=0$ has been taken from model sequence A12.  
}\label{fig:further}
\end{figure}

In an attempt to estimate $\tau_{\rm delay}$, we have selected
a white dwarf model which has grown in mass to 1.50 \Msun{} in model sequence A12, 
and let it evolve without further mass accretion. 
The evolution of the angular velocity in the white dwarf model
is shown in Fig.~\ref{fig:further}.
The initial model of this calculation is characterized by
$T_{\rm c} = 1.37 \times 10^8$ K, $\rho_{\rm c} = 1.87 \times 10^8~{\rm g/cm^3}$ 
and \TW{} = 1.11.
As the white dwarf cools, the buoyancy force in the white dwarf
interior becomes weaker and thus the dynamical shear instability 
continues to operate in the central core even though
the degree of the differential rotation becomes continuously smaller. 
Only when $t \simeq 1.1 \times 10^8$ yr, 
the shear strength becomes so weak that the shear instability finally decays. 
The central temperature and density at this moment
are $4 \times 10^7$ K and $1.57 \times 10^8~{\rm g/cm^3}$ respectively. 
This example indicates that
the CFS instability can begin to operate
about $10^{\rm 8}$ yr after the end of the mass accretion. 
Which instability mode is responsible for the following
angular momentum loss depends on the value of \TWc{} of the
white dwarf configuration at this point.

If the angular momentum loss time scale is short enough, 
(i.e., $\tau_J \lsim 10^7$),
the white dwarf may end in a Type Ia supernova,
since the central temperature is still relatively high, and 
it will increase by the contraction of the core as the white dwarf loses
angular momentum.  Piersanti et al. (\cite{Piersanti03}) 
showed that an initially cool Chandrasekhar mass white dwarf 
with $T_{\rm c} \simeq 5.6 \times 10^7 $~K
is heated to $T_{\rm c} \sim 10^8$ K  
if the white dwarf loses angular momentum
on a time scale of $10^4 \dots 10^7$ yr. 

In the example given above,
\TW{} is decreased from 0.11 to 0.10 during the evolution.
If \TWc{} at this point is less than 0.1, 
the bar-mode instability will remove angular momentum in $\tau_{J,{\rm bar}}$,
which may be short enough (i.e., $\lsim 10^7$ yr) to produce
a SNe Ia. 
On the other hand, if \TWc{} is larger than 0.1 at the given moment, 
the r-mode instability will become dominant. The value of
$\tau_{\rm r}$ at this point is about $2\times10^5$~yr.
If the time scale for the following angular momentum loss 
is of the order of $\tau_{\rm r}$, the white dwarf will safely end in a SN Ia.

Although this discussion is somewhat speculative, it may
suggest an interesting implication for the masses of white dwarfs 
in SNe Ia explosions. 
If the CFS instability is suppressed by the shear instability,  
white dwarfs can grow in principle until they become
dynamically unstable (\TW $\gsim$ 0.27) without becoming
secularly unstable. In most close binary systems,
the mass accretion will stop before reaching this point, 
given that the mass at \TW{} = 0.18 ($M_{0.18}$) in our models can be already as high as 
1.45 \Msun{} -- 1.75 \Msun{} (Table~\ref{tab:result2}).
Therefore, the upper limit for the white dwarf
mass at the moment of the supernova explosion (\Msn)
is determined by the maximum possible mass a white dwarf can 
achieve by the mass accretion, (\Mmax).   
The lower limit for \Msn{} is simply 1.4 \Msun.
I.e., the white dwarf mass at the moment
of supernova explosion will be inbetween 1.4 \Msun{} and \Mmax{} (
1.4 \Msun{} $\lsim$ \Msn{} $\lsim$ \Mmax{}). 
 
The lower limit for \Msn{} (1.4 \Msun)
can not be different for different initial masses.
However, the upper limit (\Mmax) may be systematically 
larger for higher initial masses. I.e., Langer et al. (\cite{Langer00}), 
found that \Mmax{} is larger for a higher \Minit{} in main sequence star + white 
dwarf binary systems.
Furthermore,  \Mmax{} is expected to show a dependence
on the metalicity of the mass donor.
Langer et al. (\cite{Langer00}) showed that \Mmax{} for
main sequence star + white dwarf binary systems 
decreases for lower metalicity.
This implies 
a higher probability to obtain massive exploding white dwarfs
for higher metalicity. 
If more massive white dwarfs gave brighter SNe Ia  as Fisher et al. (\cite{Fisher99}) 
speculate, the brightest SNe Ia observed at low metalicity 
may be dimmer than those at higher metalicity, while
the luminosity of the dimmest ones may not be much different (unless
affected by the CO ratio; Umeda et al.~\cite{Umeda99}, H\"oflich et al.~\cite{Hoeflich00}).

\subsubsection{Case II}

If the CFS instability is not affected by 
the shear instability, 
an accreting white dwarf will start losing angular 
momentum when $\tau_{\rm r}$ or $\tau_{\rm bar}$ 
becomes smaller than the accretion time scale 
($\tau_{\rm acc} \simeq \Delta M/\dot{M}$).

If we consider sequence A6 as an example,
\taur{} becomes comparable to \taua{} only when \TW{}~$\simeq$~0.14. 
Therefore, if we assume \TWc{} $=0.1$, the bar-mode instability
will dominate before the $r$-mode instability becomes important.
Once \TW{} becomes larger than 0.1, 
the white dwarf will lose angular momentum via the bar mode instability
while gaining angular momentum continuously from the accreted matter. 
It is likely that \TW{} will not change much  after
\taub~$\approx$~\taua{} is obtained.
This means that \TW{} can not increase much from the critical value of 0.1, 
since \taub{} drops rapidly as it deviates from \TWc{} (Fig.~\ref{fig:tau}).
If mass accretion continues, the white dwarf will finally  reach
carbon ignition at \Mwd{}~$\simeq$~1.76 \Msun, 
according to the results of sequences C9 $\dots$ C12, 
where carbon ignites at the center
when \Mwd{}~$\simeq$~1.76 \Msun{}, with \TW{}~$\simeq$~0.11 
(see also Sect.~\ref{sect:explosion}). 
If mass accretion stops before carbon ignition is reached but
after the white dwarf mass has grown beyond 1.4 \Msun, 
carbon ignition will be delayed until the white dwarf loses
enough angular momentum.

In summary, we can expect that even in Case~II 
the mass of exploding white dwarfs will show a significant diversity, as 
in Case~I.  The lower limit for \Msn{} should be $\sim$1.4 \Msun{}
in both cases, 
but the upper limit may be either the critical mass for carbon ignition
at that \TW{} which gives $\tau_{\rm J} \approx \tau_{\rm acc}$, 
or the maximum possible achievable mass by mass accretion.

\subsubsection{Other uncertainties}

In reality, the mass transfer rate in a close binary system
does not remain constant. For instance, in a binary system consisting of
a main sequence star + a white dwarf, where thermally unstable mass transfer occurs,
the mass transfer rate increases rapidly from the onset 
of the mass transfer, followed by slow decrease to the point that
the nuclear shell burning can not be stable any more (Langer et al.~\cite{Langer00}).  
Although  shell sources may be more stable with rotation (Yoon et al.~\cite{Yoon04c}),
the white dwarf will experience strong shell flashes
if the mass accretion rate decreases below $10^{-7}$ \msyr. 
This may cause a significant loss of mass, 
which may even decrease the white dwarf mass. 
This may also lead to a removal of angular momentum from the white dwarf (Livio \& Pringle~\cite{Livio98}). 
Effects of magnetic fields, if they are strongly amplified due to the differential
rotation (Spruit~\cite{Spruit02}), may also serve to break the white dwarf by 
magnetic torques and/or by magnetic dipole radiation.
These possibilities will be subject of further studies.

\section{Implications for the SNe Ia diversity}\label{sect:diversity}

We note that for all considered circumstances,
effects of rotation in accreting white dwarfs offer the possibility
of SN Ia explosions from white dwarfs more massive than $\sim$1.4~\Msun{} 
(hereafter designated as super-Chandrasekhar mass).
The observational signatures of super-Chandrasekhar mass explosions
are currently difficult to predict.
The peak brightness and light curves
of SNe Ia might be a sensitive function of the ignition conditions
such as the CO ratio, the density at the moment of carbon ignition, the speed of
the deflagration front, the core rotation rate, and possibly the transition density at which the
deflagration turns into a detonation 
(e.g. Khokholov~\cite{Khokhlov91a}, \cite{Khokhlov91b}; H\"oflich \& Khokhlov~\cite{Hoeflich96a};
Niemeyer \& Woosley~\cite{Niemeyer97}; 
H\"oflich et al.~\cite{Hoeflich98}; Iwamoto et al.~\cite{Iwamoto99}; 
Umeda et al.~\cite{Umeda99}; Hillebrandt \& Niemeyer~\cite{Hillebrandt00}; 
Woosley et al.~\cite{Woosley03}). 

%Nevertheless, it is a matter of interest how different masses of exploding
%white dwarfs are correlated to the outcome of explosion, since
%the previous discussions indicate that \Msn{} may vary
%from 1.4~\Msun{} which is the canonical Chandrasekhar limit,  
%to \Mmax{} which is the upper limit for a white dwarf to be able
%to achieve by mass accretion in a binary system unless
%angular momentum is lost during the mass accretion phase.
%According to Langer et al. (2000), \Mmax{} can 
%be, in principle, as large as 2.0~\Msun{} 
%in case \Minit{} is 1.0~\Msun, although
%accretion of such a large amount of mass
%can be achieved rather rarely, 
%compared to the cases of \Msn{}~$\approx$~1.4~\Msun.

If the ignition conditions are not strongly affected by the
mass of the exploding white dwarf, 
it is likely that more massive one gives a brighter SN Ia,
since a more massive white dwarf can provide more fuel to produce \Ni{56}.
For instance, progenitors with a super-Chandrasekhar mass
are often invoked to explain such an anomalous SN Ia as SN 1991T. 
Fisher et al. (\cite{Fisher99}) note that the mass of \Ni{56} produced 
from the explosion models  of a Chandrasekhar mass white dwarf remains to be smaller 
than about 0.9~\Msun{} even in the case
where a pure detonation is considered (e.g. H\"oflich \& Khokhlov~\cite{Hoeflich96a}). 
Although the nickel masses derived from the luminosities of normal SNe Ia
are constrained to 0.4~$\dots$~0.8~\Msun{} (Leibundgut~\cite{Leibundgut00}), 
the existing models fail
to explain such a peculiar SN Ia as 1991T, 
whose brightness implies the production of about 1.0~\Msun{} of \Ni{56}.
The fact that explosion models with the canonical Chandrasekhar mass 
could not explain the peculiarity of SN 1991T suggests that it is worthwhile
to investigate the possible outcome of super-Chandrasekhar mass
explosions,
even if we can not exclude the possibility that 
the luminosity of 1991T is biased by the uncertainty
in determining the distance of its host galaxy (Hanato et al.~\cite{Hanato02}).

On the other hand, Fisher et al. (\cite{Fisher99}) suggested 
the explosion of a super-Chandrasekhar mass white dwarf 
from the merger of double CO white dwarfs as explanation of
the overluminosity found in SN 1991T. 
We  note that differentially rotating single degenerate progenitors
may be a more natural explanation for the super-Chandrasekhar mass scenario
than double degenerate mergers, which fail to produce SNe Ia in 
numerical models due to the off-center carbon ignition 
induced by the fast mass accretion with \Mdot{}~$\gsim 10^{-5}$~\msyr{} 
(Saio \& Nomoto~\cite{Saio85},~\cite{Saio98}).

Rotation also bears implications for the polarization of SNe~Ia.
According to our results, 
the rotation velocity
of exploding white dwarfs may strongly depend
on the history of angular momentum loss
via gravitational wave radiation.
The more angular momentum is lost, the slower
the white dwarf rotation will be, 
which means that the polarization strength in SNe Ia 
may show a significant diversity.
Some white dwarfs may have a chance to end in an SN Ia 
while they are rotating very rapidly,
as in sequences C9, C10, C11 and C12, 
where carbon ignition occurs at the white dwarf center
when \Mwd{}~$\simeq$~1.76 \Msun{} and \TW{}~$\simeq$~0.11. 
Such exploding white dwarfs 
will show strong features of asphericity in 
the explosion, 
which might give a plausible explanation for 
the polarization observations in SN 1999by  (Howell et al.~\cite{Howell01})
and SN 2001 el (Wang et al.~\cite{Wang03}; Kasen et al.~\cite{Kasen03}).

\section{Pre-explosion conditions of a fast rotating white dwarf}\label{sect:explosion}

\begin{figure}
\resizebox{\hsize}{!}{\includegraphics{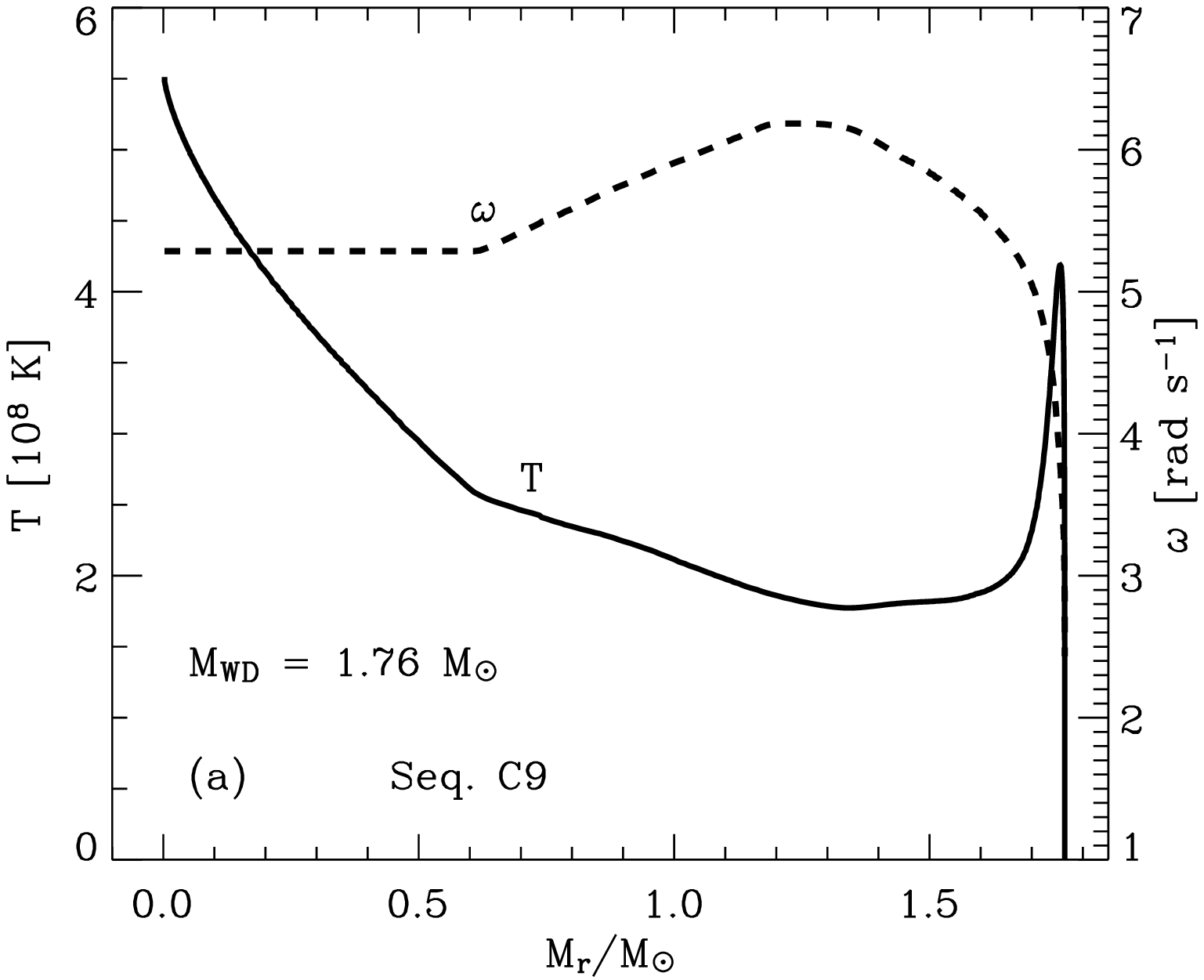}}
\resizebox{\hsize}{!}{\includegraphics{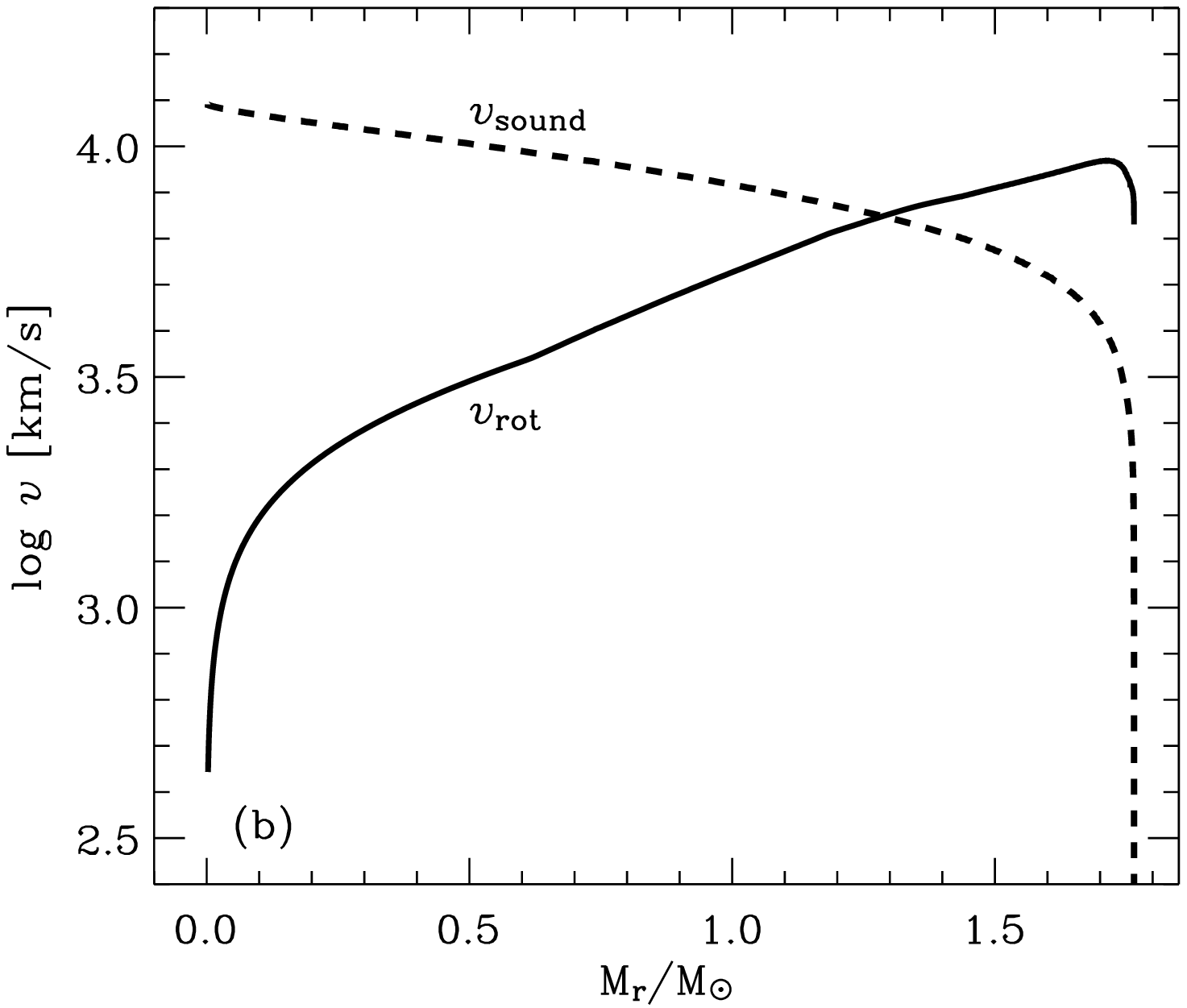}}
\caption{ (a) Temperature (solid line) and angular velocity (dashed line) as function 
of the mass coordinate in the last computed model of sequence C9. 
(b) Equatorial rotational velocity (solid line) for the same model as in (a) as function
of the mass coordinate. 
The dashed line denotes the sound speed as function of the mass coordinate.
}\label{fig:prexp}
\end{figure}

Rotation may have interesting consequences for the supernova explosion
itself.
Fig.~\ref{fig:prexp} shows the physical properties of the white dwarf model
of sequence C9 when \Mwd{}~=~1.76~\Msun, 
where the central temperature and density reach  $5.5 \times 10^8$~K
and $2.2\times10^9$~\density, respectively.
The central region ($0.0 \lsim M_{\rm r} \lsim 0.6 {\rm M}_{\odot}$) 
is found convectively unstable due to carbon burning, 
rotating rigidly
due to the convective angular momentum redistribution. 
The thermonuclear runaway is expected to develop when 
the central temperature reaches a few times $10^9$~K. 
The convective core will be more extended by then,
spinning-up the central region further.
In the model shown in Fig.~\ref{fig:prexp}, 
the central region rotates with $v_{\rm rot} \simeq 1000$~km/s, 
which is well above the convective velocity 
($\sim 100$~km/s; e.g. H\"oflich \& Stein~\cite{Hoeflich02}) 
and somewhat larger than the expected initial velocity of the deflagration
front 
($v_{\rm def,init} \simeq 0.03 v_{\rm sound}$, e.g. Nomoto et al.~\cite{Nomoto84}). 
The equatorial rotation velocity increases sharply 
from the center to the edge of the convective layer ($M_{\rm r} = 0.6$~\Msun),
by a factor of 10.

This fast rotation in the central carbon burning core may affect
the explosion as follows. 
Numerous studies indicate that
the combustion front develops large scale turbulent motions, especially due to the Rayleigh-Taylor
instability (e.g. Reinecke et al.~\cite{Reinecke02}; Gamezo et al.~\cite{Gamezo03} and references therein). 
It is a crucial question whether this turbulence can induce a detonation 
(Khokhlov~\cite{Khokhlov91a},~\cite{Khokhlov91b};
Niemeyer \& Woosley~\cite{Niemeyer97}; Khokhlov et al.~\cite{Khokhlov97}). Although pure deflagration
models show a good agreement with observations 
(Nomoto et al.~\cite{Nomoto84}; H\"oflich \& Khokhlov~\cite{Hoeflich96a}),
recent three dimensional calculations show that large amounts of unburned and partially 
burnt materials are left even at the center (Reinecke et al.~\cite{Reinecke02}; Gamezo et al.~\cite{Gamezo03}), 
which is not supported by observations (however, see Baron et al.~\cite{Baron03}). 
This problem would disappear if a detonation were triggered by the turbulent deflagration 
('delayed detonation', Khokhlov~\cite{Khokhlov91a}). 
Furthermore, delayed detonation models are shown to reproduce
the light curves and spectra of well observed Type~Ia SNe
(H\"oflich \& Khokhlov~\cite{Hoeflich96a}; H\"oflich et al.~\cite{Hoeflich96b}). 
A study of the nucleosynthesis output of SNe~Ia by Iwamoto et al. (\cite{Iwamoto99}) also favors 
the delayed detonation scenario.

Several authors conclude that the probability for triggering a detonation 
can be enhanced if the turbulence
is strong enough to form a thick mixed region of 
unburned and burning materials
with a constant temperature gradient 
(Khokhlov~\cite{Khokhlov91b}; Niemeyer \& Woosley~\cite{Niemeyer97}; Khokhlov et al.~\cite{Khokhlov97}; 
Lisewski et al.~\cite{Lisenwski00}). 
However, no robust mechanism for the formation of such a strong turbulence 
has yet been suggested.
(Niemeyer~\cite{Niemeyer99}).

A rapidly rotating white dwarf like the one shown in Fig.~\ref{fig:prexp}
will be spun down due to expansion once the thermal runaway occurs,
as the deflagration wave propagates.
However, the outer layers will still rotate rapidly at $v_{\rm rot}\sim 10^3$~km/s.
Lisenwski et al. (\cite{Lisenwski00}) show that strong turbulence, i.e. 
$v_{\rm turb} \approx 10^3$ km/s, is necessary for a detonation to form successfully.
Since the rotation velocity in the outer region will be of this order,
the shear motion may provide enough kinetic energy for the
turbulence intensity to satisfy the condition for a transition
of the deflagration into a detonation,
which might not be possible in the non-rotating case since 
the maximum value for $v_{\rm turb}$ is $\sim 10^2$ km/s in the Rayleigh-Taylor-driven
turbulence.  Alternatively, the fast rotation in the outer region 
can tear the turbulent deflagration front and enhance
the nuclear burning surface, which may be another way to trigger a detonation
(H\"oflich~2003, private communication).
Future multi-dimensional calculations are required to 
test the validity of these scenarios.

\section{Gravitational wave radiation in pre-explosion white dwarfs?}\label{sect:gwr}

Rotation may have another important and exciting observational consequence. 
Our results indicate
that the white dwarfs in SN~Ia progenitor systems may
emit gravitational
wave radiation (GWR). Therefore, if we could observe
GWR preceding a SN Ia event, 
this would provide strong evidence for the scenarios outlined
in Sect.~\ref{sect:final}.

The GWR due to the CFS bar-mode instability may
be much stronger than the r-mode signal, since
perturbations due to the bar-mode involve larger density
changes than those due to the r-mode (e.g. Andersson \&  Kokkotas~\cite{Andersson01}).
Fryer et al. (\cite{Fryer02}) derive
the gravitational wave amplitude observed at Earth from a source which undergoes the bar-mode instability  
at a distance $d$ as
\begin{equation}
h_{\rm bar} = \left(\frac{32}{45}\right)^{1/2} \frac{G}{c^4}\frac{MR^2\Omega^2}{d} ,
\end{equation}
where $R$ denotes the equatorial radius of the source, $M$ the mass and $\Omega$  the angular velocity.
The frequency of the gravitational waves ($f_{\rm GW}$) due to the bar mode instability  
is twice the rotation frequency of the source. 
We estimate $h_{\rm bar}$ and $f_{\rm GW}$ 
for selected white dwarf models in Table~\ref{tab:gwr}. 
For these calculations, we used mean values for $\Omega$ and $R$
given in Table~\ref{tab:result1} and~\ref{tab:result2}. 

\begin{table}[t]
\begin{center}
\caption{Expected properties of the gravitational waves from the selected accreting white dwarfs. 
}\label{tab:gwr}
\begin{tabular}{r c c c c }
\hline \hline
 No. & $M$   & \TW  & $h_{\rm bar}$ at 10 Mpc  & $f_{\rm GW}$ \\
     & \Msun &      &    $10^{-24}$             &  Hz  \\     
\hline

 A2  & 1.18  & 0.10 &   1.7    & 0.22                \\
 A6  & 1.30  & 0.10 &   2.9    & 0.32                \\
 A10 & 1.42  & 0.10 &   4.3    & 0.46                \\
 A2  & 1.34  & 0.14 &   3.7    & 0.32                \\
 A6  & 1.44  & 0.14 &   5.0    & 0.43                \\ 
 A10 & 1.55  & 0.14 &   7.5    & 0.62                \\
\hline
\end{tabular}
\end{center}
\end{table}

Table~\ref{tab:gwr} indicates that $f_{\rm GW}$
is within the range of $0.1 \dots 1.0$~Hz.
Observation of the GWs with such frequencies  
may be performed by low frequency
gravitational wave detectors such as LISA,  which
will cover the range $10^{-4} - 1.0$~Hz (e.g. Cutler \& Thorne~\cite{Cutler02}).
The expected strength of the gravitational wave signal is
about $10^{-24}$ at $d=10$~Mpc. 
Hiscock (\cite{Hiscock98}) estimated the strength of GWs due to
the $r$-mode instability from white dwarfs 
in the observed DQ Her systems as $h\sim 10^{-23}$.
His calculations show that this strength is well above
the detection limit of the LISA interferometer. 
Therefore, the results shown in Table~\ref{tab:gwr} imply
that SNe Ia progenitors in nearby galaxies
could be within the observable range. 
Given that GWs from SN~Ia progenitors
will be emitted over a secular time, and that
SNe Ia are observed with a relatively high rate 
of $ \sim 3 \times 10^{-3}~{\rm yr^{-1}}$ per galaxy, 
the probability to detect the GW signal may be significant.

Even if a white dwarf would not reach
the bar-mode instability at \TWc{},
it might still emit gravitational waves via the $r$-mode
instability. Although their strength is rather unclear and
may be much weaker than that of the bar-mode,
they may still be detectable if the source is close enough.

\section{Conclusion}\label{sect:conclusion}

We summarize the results of this paper as follows.

1. The role of the Eddington sweet circulation, the GSF instability 
and the shear instability for the transport of angular momentum 
in non-magnetized white dwarfs has been investigated
(Sect.~\ref{sect:angmom}).
Although Eddington sweet circulation and the GSF instability
are important for the redistribution of angular momentum
in the non-degenerate envelope, their importance
is small in the degenerate core compared to the shear instability.
The secular shear instability 
can not operate in the strongly degenerate core, since 
the thermal diffusion time becomes longer
than the turbulent viscous time for 
densities higher than a critical value (i.e., 
$\rho \gsim 10^6 \dots 10^7 ~{\rm g/cm^3}$, Fig.~\ref{fig:rho_ssi}).
On the other hand, the criterion for the dynamical shear instability
is significantly relaxed for higher density 
because the buoyancy force becomes weaker with stronger degeneracy (Fig.~\ref{fig:sigma}).
As a result, the degenerate inner core is dominated by 
the dynamical shear instability in accreting white dwarfs as shown in Sect.~\ref{sect:spin}. 

2. We have followed the redistribution of the angular momentum
in accreting white dwarfs by the above mentioned processes 
(Sect.~\ref{sect:spin}).
We find that accreting white dwarfs do not rotate 
rigidly, but differentially throughout their evolution, for the 
considered accretion rates (\Mdot{} = $3\dots10 \times 10^{-7}$ \msyr).
In the degenerate core, once the shear factor
decreased to the threshold value for the onset of the 
dynamical shear instability, the time scale for further angular momentum
transport becomes larger than the accretion time scale.
Accordingly, 
strong differential rotation is retained in the inner core
with a shear strength near the threshold value for the dynamical shear
instability (Fig.~\ref{fig:sigma_6a}).

3. Accreting white dwarfs, as they rotate differentially, may not reach
central carbon ignition even when they grow beyond the
canonical Chandrasekhar limit of $\sim$1.4~\Msun. 
This is in accordance with previously obtained 
results based on other methods
(Ostriker \& Bodenheimer~\cite{Ostriker68a},~\cite{Ostriker73}; Durisen~\cite{Durisen75b}; 
Durisen \& Imamura~\cite{Durisen81}). 
A secular instability to gravitational wave radiation through
the $r$-mode or the bar-mode may be important for determining the final fate
of accreting white dwarfs. 
The masses of exploding white dwarfs are expected to vary in the range from the 
canonical Chandrasekhar mass ($\sim$1.4 \Msun) to the maximum possible mass 
that the white dwarf can achieve by mass accretion in a binary system 
($\sim 2.0$~\Msun, Langer et al.~\cite{Langer00}). 
This may have consequences for the diversity 
in the brightness and polarization of SNe~Ia.

4. Fast rotation in the white dwarf core may change the supernova explosion
since it can affect the evolution of the turbulent nuclear burning flames by providing
a large amount of turbulent kinetic energy and/or by enhancing the burning surface 
significantly. It needs to be clarified 
whether this may be a plausible mechanism to induce
the transition from deflagration to detonation.

5. White dwarfs which accreted enough angular momentum 
may be detectable sources of gravitational waves in the near future. 
Our models show that 
these will emit gravitational waves with frequencies of $0.1 - 1.0$ Hz. 
Space-based interferometric gravitational wave detectors such as LISA
could observe such signals from rapidly rotating SNe Ia progenitors in nearby galaxies.

\begin{acknowledgements}
We are grateful to Axel Bona$\check{\rm c}$i\'c and Philipp Podsiadlowski for
many useful discussions.
This research has been supported in part by the Netherlands Organization for
Scientific Research (NWO). 

\end{acknowledgements}

\end{document}